\documentclass[aps,pra,superscriptaddress,amsmath,amssymb,floatfix,twocolumn]{revtex4}
\usepackage{bbold}
\usepackage{graphicx}
\usepackage{subfigure}
\usepackage{color}
\usepackage{hyperref}
\usepackage{comment}

\begin{document}

\title{Dynamic zero modes of Dirac fermions and competing singlet phases of antiferromagnetic order}

\author{Pallab Goswami}
\affiliation{Condensed Matter Theory Center and Joint Quantum Institute, Department of Physics, University of Maryland, College Park, Maryland 20742- 4111 USA}
\author{Qimiao Si}
\affiliation{Department of Physics and Astronomy, Rice University, Houston, TX 77005, USA}

\begin{abstract}
In quantum spin systems, singlet phases often develop in the vicinity of an antiferromagnetic order. 
Typical settings for such problems arise when itinerant fermions are also present.
In this work, we develop a theoretical framework for addressing such competing orders in an itinerant system, described by Dirac fermions strongly coupled to an O(3) nonlinear sigma model. We focus on two spatial dimensions, where upon disordering the antiferromagnetic order by quantum fluctuations the singular tunneling events also known as (anti)hedgehogs can nucleate competing singlet orders in the paramagnetic phase. In the presence of an isolated hedgehog configuration of the nonlinear sigma model field, we show that the fermion determinant vanishes as the dynamic Euclidean Dirac operator supports fermion zero modes of definite chirality. This provides a topological mechanism for suppressing the tunneling events. Using the methodology of quantum chromodynamics, we evaluate the fermion determinant in the close proximity of magnetic quantum phase transition, when the antiferromagnetic order parameter field can be described by a dilute gas of hedgehogs and antihedgehogs. We show how the precise nature of emergent singlet order is determined by the overlap between dynamic fermion zero modes of opposite chirality, localized on the hedgehogs and antihedgehogs. For a Kondo-Heisenberg model on the honeycomb lattice, we demonstrate the competition between spin Peierls order and Kondo singlet formation, thereby elucidating its global phase diagram. We also discuss other physical problems that can be addressed within this general framework.
  
\end{abstract}

\maketitle

\section{Introduction}
The competition 
between spin-singlet phases and antiferromagnetic order is a common feature of the phase diagrams for many strongly correlated systems, such as heavy fermion compounds, cuprates, 
and
iron pnictides. Depending on the context, the singlet order can correspond to unconventional superconductivity, charge, bond and current density waves, and static Kondo singlets. The competition between singlet and triplet orders can cause an exotic quantum critical point or an intervening non-Fermi liquid phase between two  distinct broken symmetry states. A prototype case arises in heavy fermion metals, where non-Fermi liquid properties 
arise in the quantum critical regime~\cite{Si-Nature,Coleman-JPCM,Senthiletal,Paschen,Shishido,Si_PhysicaB2006,Lohneysen_rmp,SiSteglich}. 
The latter is typically associated with a competition between the antiferromagnetic order of the local moments and the Kondo-singlet
 or related
phases,
and a global phase diagram
has been advanced to capture the variety of spin-singlet phases near the antiferromagnetic order~\cite{Si_PhysicaB2006}. 
Therefore, it is imperative to develop a general scheme for identifying competing singlet orders beginning from the magnetically ordered phase and vice versa. 

\begin{figure}[htb]
\begin{center}
\includegraphics[scale=0.6]{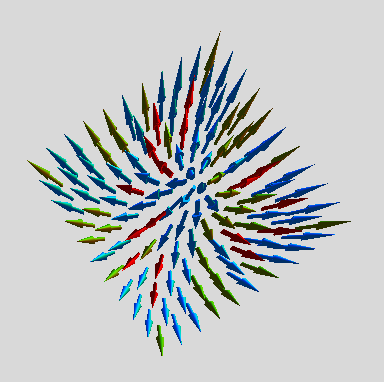}
\caption{Illustration of a skyrmion texture of nonlinear sigma model field with topological invariant $+1$ in the $x-y$ plane. The skyrmion texture is well defined only inside the magnetically ordered phase, when the tunneling singularities shown in Fig.~\ref{fig2}, namely the hedgehog and antihedgehog remain linearly confined. The skyrmion core can support several fluctuating translational symmetry breaking or intervalley orders for a single species of Dirac fermion. For two or more species of Dirac fermions it can also support translational symmetry preserving inter-species orders such as Kondo singlets.}
\label{fig1}
\end{center}
\end{figure}

In this paper, we will develop such a scheme for an itinerant system of (2+1) dimensional massless Dirac fermions which are strongly coupled to an O(3) nonlinear sigma model, by considering the interplay between fermionic degrees of freedom and the topological defects of antiferromagnetic order parameter. Inside the antiferromagnetically ordered phase, the low energy spin wave excitations or Goldstone modes are well described by an $O(3)$ nonlinear sigma model~\cite{Duncan,Chakravarty,MurthySachdev,Read}. However, it is tailored for capturing the smooth collective excitations, and will be impervious to the presence of competing singlet orders. For insulating systems described by generalized Heisenberg models, it has been proposed that the topological excitations of (2+1) dimensional nonlinear sigma model can give rise to singlet valence bond solid order in the paramagnetic phase~\cite{Duncan,Read}. On the magnetically ordered side, competing singlet orders can reside inside the core of a topological but nonsingular skyrmion defect (see Fig.~\ref{fig1}). Since the skyrmion is a finite energy excitation, the singlet orders can exist only as gapped, fluctuating quantities in the magnetically ordered phase. By contrast, due to the vanishing of spin stiffness inside the paramagnetic phase, the skyrmion excitation gap disappears, causing an enormous degeneracy among topologically distinct ground states, labeled by different skyrmion numbers. Therefore, inside the paramagnetic phase different types of singlet orders can be nucleated by breaking the skyrmion number conservation. 

\begin{figure}[htbp]
\centering
\subfigure[]{
\includegraphics[scale=0.55]{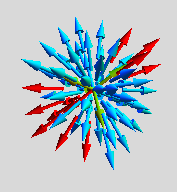}
\label{fig2a}
}
\subfigure[]{
\includegraphics[scale=0.53]{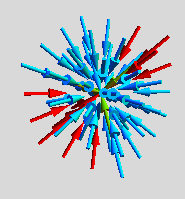}
\label{fig2b}
}
\caption[]{Illustration of nonlinear sigma model field around (a) a radial hedgehog and (b) a radial antihedgehog singularities in Euclidean space-time. They represent tunneling events which change the skyrmion number of the background by one. On the magnetic side they stay linearly confined without influencing the low energy physics. However on the paramagnetic side, they are no longer confined and play dominant role in determining the nature of emergent competing order. Overall neutrality condition for the background field requires an equal number of hedgehog and antihedgehog singularities on average, and we show one pair of these singularities in Fig.~\ref{fig3}.}\label{fig2}
\end{figure}

The huge ground state degeneracy of the paramagnetic phase is generally lifted by the topological hedgehog singularities (see Fig.~\ref{fig2}), which describe tunneling between ground states with different skyrmion numbers. The hedgehogs are generally accompanied by a dynamic Berry phase term for the nonlinear sigma model~\cite{Duncan,Read}. Within the coarse grained description of magnetic order, the Berry phase carries important information regarding the quantized value of microscopic spin and determines the nature (or reduced degeneracy) of singlet order in the paramagnetic phase. Based on this physical picture, an exotic continuous quantum phase transition between two distinct broken symmetry phases has been proposed~\cite{Senthiletal2,Senthiletal3}, which falls outside the paradigm of conventional Landau-Ginzburg theory. Within the CP$^1$ formulation of the sigma model, both magnetic (Higgs phase of gauge theory) and paramagnetic spin Peierls phases are confined states of an underlying compact U(1) gauge theory. It has been suggested that only at the critical point the hedgehogs or monnopoles can be suppressed, leading to the deconfined or noncompact U(1) gauge field and spinon excitations. There are ongoing numerical studies on different microscopic models which are providing encouraging evidence for an exotic direct transition between two ordered states~\cite{Sandvik,Kaul,Chalker1,Chalker2}. However, numerical works also find strong violations of hyperscaling and the issue of deconfined criticality is not yet settled.

How does the coupling between antiferromagnetic order parameter and itinerant fermions affect this scenario? In the present work, we will be addressing this important question. We want to compute the fermion determinant for the following model 
\begin{equation}
S_1=\int d^3x\;  \bar{\psi} \left[\sum_{\mu=0}^{2} \; \Gamma_\mu \otimes \sigma_0 \partial_\mu - i m \; \Gamma_3 \otimes \boldsymbol \sigma \cdot \mathbf{n} \right]\psi,   \label{s1}
\end{equation}
and its suitable generalizations, where $\psi$ is an eight component Dirac spinor comprising of two sublattices, two valleys and two spin components, $\Gamma_\mu$ and $\Gamma_3$ are $4\times 4$ Hermitian anticommuting matrices, and $\sigma_j$s are Pauli matrices operating in the spin space. The nonlinear sigma model field is described by the unit vector $\mathbf{n}$ and $m$ is a coupling constant. This  model will be generally augmented by a standard effective action for nonlinear sigma model. But, in this work we are mainly  interested in computing the fermionic contribution to the effective action in the paramagnetic phase, when $\mathbf{n}$ only displays short range correlations. We note that the dynamic Dirac operator anticommutes with the fifth gamma matrix $\Gamma_5$. For convenience, we will work with a block off-diagonal representation of the Dirac operator with the following choice of gamma matrices: $\Gamma_0=\eta_1 \otimes \tau_3$, $\Gamma_1=\eta_1 \otimes \tau_1$, $\Gamma_2=\eta_1 \otimes \tau_2$, $\Gamma_3=\eta_2 \otimes \tau_0$ and $\Gamma_5=\eta_3 \otimes \tau_0$, where Pauli matrices $\eta_\mu$s and $\tau_\mu$s respectively operate on valley and sublattice indices.  

\subsection{Nucleation of spin Peierls order}
From previous perturbative calculations (gradient expansion controlled by the local gap $m$) on the magnetically ordered side~\cite{Wilczek1,Jaroszewicz,Wilczek2,Abanov,Hermele,FisherSenthil,TanakaHu,FuSachdev,RoyHerbut,GoswamiSi2} we know that the skyrmion core can support several translational symmetry breaking (mixing two valleys) charge, bond and current density wave orders. Their explicit forms are given by the pairs
$$(i) \; (\bar{\psi}\psi, i\bar{\psi}\Gamma_5\psi), \; (ii) \; (\bar{\psi}\Gamma_{0j}\psi, i\bar{\psi}\Gamma_5\Gamma_{0j}\psi),$$ with $j=1,2,3$. 
While the Dirac mass terms $(\bar{\psi}\psi, i\bar{\psi}\Gamma_5\psi)$ describe spin Peierls order, the other three pairs (not mass terms) correspond to charge and current density wave orders. For the Dirac fermions obtained from the honeycomb lattice, $j=1,2,3$ pairs respectively describe (a) intervalley, nonstaggered, intrasublattice charge (b) intervalley, staggered intrasublattice charge, (c) intervalley, intersublattice current density waves. \emph{What is the explicit mechanism by which the system determines the form of nucleated singlet order in the paramagnetic phase, where skyrmion number is no longer conserved?} 

\begin{figure}[htb]
\begin{center}
\includegraphics[scale=0.6]{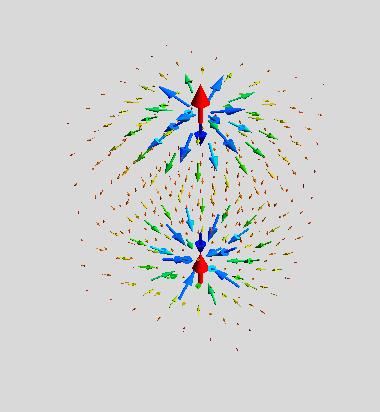}
\caption{A pair of hedgehog and antihedgehog separated by a distance $R$ along imaginary time direction. Since hedgehog and antihedgehog are respectively the monopole and antimonopole of $CP^1$ gauge field, there will be quantized $2\pi$ amount of $CP^1$ flux through the $xy$ plane, which is perpendicular to the direction of their separation. The situation is analogous to how $2\pi$ Berry flux passes through a plane perpendicular to the separation vector of left and right handed Weyl points in a three dimensional Weyl semimetal. On the paramagnetic side, but close to the magnetic quantum critical point, the separation $R$ varies as $\xi^x$, where $\xi$ is the correlation length of nonlinear sigma model field. In the vicinity of quantum critical point, the diverging $\xi$ implies a very large separation $R$, which allows us to perform calculations with a dilute hedgehog gas approximation.}
\label{fig3}
\end{center}
\end{figure}

We show that in the presence of an isolated hedgehog singularity $\mathbf{n}= \hat{x}$ (see Fig.~\ref{fig2}), the Euclidean Dirac operator $$\mathcal{D}=\Gamma_\mu \partial_\mu - i m \Gamma_3 \otimes \boldsymbol \sigma \cdot \mathbf{n}$$ supports two fermion zero modes of positive chirality (valley index= $+1$) with an antisymmetric locking of spin and sublattice indices ($\epsilon_{a\alpha}$). By contrast, two zero modes for the antihedhehog singularity possess opposite chirality (valley index $-1$), but the same form of spin-sublattice locking. The source of opposite chirality is the spectral symmetry condition $\{\mathcal{D},\Gamma_5\}=0$, and the zero modes are protected by Callias index theorem~\cite{JackiwRebbi1,Callias1,Callias2}. \emph{The fermion determinant vanishes due to these dynamic fermion zero modes, leading to an infinite action for an isolated tunneling event. Consequently, the probability or fugacity of a single (anti)hedgehog vanishes, providing a topological mechanism for suppressing tunneling events}. In the close proximity of magnetic quantum phase transition, we can model the neutral background field $\mathbf{n}$ in terms of a \emph{dilute gas consisting of equal numbers of hedgehogs and antihedgehogs on average} (as their mean separation is controlled by the diverging correlation length), and the overlap between localized zero modes of opposite chirality determines the precise nature of singlet order. A pair of hedgehog and antihedgehog are shown in Fig.~\ref{fig3}. \emph{By averaging over hedgehog location and orientation within the O(3) group, we unambiguously show that the resulting singlet order parameter corresponds to the dynamic, complex Dirac mass $\bar{\psi} e^{i \theta \Gamma_5} \psi$ describing the translational symmetry breaking spin Peierls order}. This method also provides valuable insight regarding the unconventional nature of quantum phase transition between two different broken symmetry phases.

\subsection{Competition between spin Peierls and Kondo singlets}
Next we consider two species of eight-component Dirac fermions $\psi$ and $\chi$ with an opposite sign of Yukawa couplings ($m_1=-m_2=m$), 
\begin{eqnarray}
S_2&=&\int d^3x\;  \bar{\psi} \left[\sum_{\mu=0}^{2} \; \Gamma_\mu \partial_\mu - i m \; \Gamma_3 \otimes \boldsymbol \sigma \cdot \mathbf{n} \right]\psi  \nonumber \\
&+&\int d^3x\; \bar{\chi} \left[\sum_{\mu=0}^{2} \; \Gamma_\mu \partial_\mu +i m \; \Gamma_3 \otimes \boldsymbol \sigma \cdot \mathbf{n} \right]\chi.   \label{s2}
\end{eqnarray}
This type of effective theory can describe the antiferromagnetic insulator phase of a Kondo-Heisenberg model on a honeycomb lattice~\cite{GoswamiSi2}. Within the gradient expansion scheme for magnetically ordered phase, previously we have identified the following translational symmetry preserving Kondo (interspecies) singlet bilinears in the skyrmion core
$$\bar{\Psi} \Gamma_\rho  \otimes \mu_{1}\Psi, \; \bar{\Psi}\Gamma_\rho \Gamma_5\otimes \mu_{1}\Psi, \; \bar{\Psi} \Gamma_\rho  \otimes \mu_{2}\Psi, \; \bar{\Psi}\Gamma_\rho \Gamma_5\otimes \mu_{2}\Psi,$$
with $\rho=0,1,2,3$, where $\Psi^T=(\psi^T,\chi^T)$ is a sixteen component spinor and Pauli matrices $\mu_j$s operate on species index~\cite{GoswamiSi2}. Additionally a skyrmion core can support the following intraspecies, translational symmetry breaking bilinears 
$$\bar{\Psi} \mu_{0/3}\Psi, i\bar{\Psi}\Gamma_5 \otimes \mu_{0,3}\Psi, \bar{\Psi}\Gamma_{0j}\otimes \mu_{0/3}\Psi, i\bar{\Psi}\Gamma_5\Gamma_{0j} \otimes \mu_{0/3}\Psi,$$ and $\mu_3$ describes species-staggering. In the paramagnetic phase, we find two additional zero modes for $\chi$ fermions. The zero modes for $\psi$ and $\chi$ fermions are of opposite chirality. \emph{Since a hedgehog altogether leads to four zero modes, after averaging over the ensemble of hedgehogs and antihedgehogs we obtain a special form of quartic interaction within the zero mode subspace, which describes the strong competition between spin Peierls order and Kondo singlets (inter species singlets)}. In particular for the above effective model, we find equally strong attractive interactions for $\bar{\Psi}\Gamma_3 \otimes \mu_{1/2}\Psi$ type Kondo singlets and $\bar{\Psi} \mu_{0}\Psi, i\bar{\Psi}\Gamma_5 \otimes \mu_{0}\Psi$ type spin Peierls bilinears. We also note that the relationship between the instantons of nonlinear sigma model and Kondo singlet formation has been discussed for (1+1)-dimensional models~\cite{Tsvelik,GoswamiSi1}. Next we consider the similarity of our (2+1)-dimensional models and (3+1)-dimensional quantum chromodynamics (QCD$_4$).

\subsection{General theme of chiral symmetry breaking by instantons}
In the absence of fermionic matter (or quarks), the tunneling events (instantons) lift the huge ground state degeneracy of topologically distinct pure gauge configurations (vacua) of the Yang-Mills field~\cite{'T Hooft,JackiwRebbi2,Shuryak,Diakonov}. When massless quarks (Dirac fermions) are coupled to a nonabelian gauge field, the (3+1) dimensional dynamic Dirac operator supports zero modes of definite chirality (left or right handed) in the presence of an instanton background~\cite{'T Hooft}. This causes a vanishing fermion determinant, and the effective action for an isolated tunneling event becomes infinite. Consequently, the probability of an isolated instanton vanishes.  The presence of fermion zero mode of definite chirality provides a topological mechanism for breaking separate number conservation laws for right and left handed fermions, a phenomenon known as the axial anomaly. For a topologically trivial gauge field configuration, comprising of equal numbers of instantons and antiinstantons, the zero modes of opposite chirality (respectively localized on the instantons and the antiinstantons) can overlap. This overlap or hybridization of fermion zero modes with opposite chirality gives rise to a dynamic mass for the quarks, describing the breakdown of chiral symmetry~\cite{'T Hooft,JackiwRebbi2,Shuryak,Diakonov}. Depending on the number of flavors $N_f$, the overlap of zero modes leads to $2N_f$ fermion interaction vertex also known as t'Hooft vertex. For $N_f=1$ it corresponds to Dirac mass. For $N_f=2$ one obtains a quartic interaction very similar to the Nambu-Jona-Lasinio model, which describes spontaneous breaking of flavor chiral symmetry. As emphasized in the previous subsections, the dynamic mass for one species of eight component Dirac fermion arising due to the hybridization of two valleys also causes a breakdown of U(1) chiral symmetry (separate number conservation laws for two valleys), which is a continuum description of discrete translational symmetry. For our $N_f=2$ model, the quartic interaction describes the competition between Kondo singlet and spin Peierls orders, which capture general forms of flavor chiral symmetry breaking. Given these similarities, we will closely follow the methodology of QCD$_4$ as described in the review works of Refs.~\onlinecite{Shuryak,Diakonov}, and also denote the hedgehog induced interaction vertex as the 't Hooft vertex. The instanton induced chiral symmetry breaking is quite general~\cite{Witten}, and can arise in many strongly interacting systems in different spatial dimensions. We will later discuss an example of a quantum spin Hall system in (2+1) dimensions, where the instantons lead to superconducting phase on the quantum disordered side~\cite{Grover,ChamonRyu,Herbut1,Moon,Chakravarty1}, as well as some higher dimensional models where $D$-dimensional Dirac fermions are coupled to a quantum disordered $O(D)$ nonlinear sigma model~\cite{Abanov}.      

Our paper is organized as follows. In Sec.~\ref{model}, we consider the continuum limit of a Kondo-Heisenberg model on a honeycomb lattice to show how the effective actions of Eq.~(\ref{s1}) and Eq.~(\ref{s2}) can arise. The topological defects of O(3) nonlinear sigma model are discussed in Sec.~\ref{defects}. In Sec.~\ref{Induced1} we obtain the induced fermion numbers of skyrmion textures for a single species of eight component Dirac fermion of Eq.~(\ref{s1}) and the list of competing singlet orders in particle-hole channel. The necessity of dynamic fermion zero modes as suggested by the gradient expansion calculations and the relation with Callias index theorem are discussed in Sec.~\ref{index}. We show the explicit form of zero mode solutions in Sec.~\ref{zeromodes}. In Sec.~\ref{overlap}, we determine the overlap between the zero modes of opposite chirality and establish the explicit mechanism behind nucleating spin-Peierls order for eight component fermions of Eq.~(\ref{s1}). In Sec.~\ref{Kondo} we consider the case of Kondo singlet formation for two species of fermions described by Eq.~(\ref{s2}), and derive an effective four fermion interaction or 't Hooft vertex within the zero mode subspace.  We show how such an interaction governs the competition between the Kondo singlet formation and spin Peierls order. The applications of our methodology for diverse problems are discussed in Sec.~\ref{applications}. We summarize our main findings in Sec.~\ref{conclusions}. The detailed derivation of 't Hooft vertex is relegated to Appendix~\ref{KondoPeierls}.

\section{Model and continuum limit}\label{model}

We focus on the following Kondo Heisenberg model on the honeycomb lattice at half filling
\begin{eqnarray}
H_2=\sum_{\langle ij \rangle}\bigg[-t \; c_{i,\alpha}^{\dagger}c_{j,\alpha}+h.c. +J_H \; \mathbf{s}_{i} \cdot \mathbf{s}_{j}\bigg] \nonumber \\ +\frac{J_{K}}{2} \; \sum_i \; c_{i,\alpha}^{\dagger} \; \boldsymbol \sigma_{\alpha \beta} \; c_{i+1, \beta} \cdot \mathbf{s}_{i},
\end{eqnarray}
where $\langle ij \rangle$ represents the pair of nearest neighbors located on two different sublattices. In the absence of Kondo coupling, the conduction and the valence bands touch at the corners of the hexagonal Brillouin zone, and possess linear dispersion in the vicinity of these points. For each spin component, this is the touching between two nondegenerate bands, which gives rise to two component Weyl fermions as low energy excitations. When we linearize the spectrum around two such inequivalent points located at $\mathbf{K}$ and $\mathbf{K}^\prime$, we arrive at the following effective action at each valley
\begin{eqnarray}
S_+=\int d^2x d \tau R^\dagger_{a}\left[\partial_\tau+i v\tau_j \partial_j\right]R_{a}, \\
S_-=\int d^2x d \tau L^\dagger_{a}\left[\partial_\tau-i v\tau_j \partial_j\right]L_{a}.
\end{eqnarray}
In the above equations $a$ is the index for the spin components, which can be thought as flavor degrees of freedom, and Pauli matrices $\tau_j$ operate on the sublattice sector. We have denoted the two component spinors around the diabolic points as $R$ and $L$, and their explicit forms are given by
\begin{eqnarray}
R^T=\left[c_{A,\alpha}(\mathbf{k}+\mathbf{K}),c_{B,\alpha}(\mathbf{k}+\mathbf{K})\right], \\ L^T=\left[c_{B,\alpha}(\mathbf{k}+\mathbf{K}^\prime),c_{A,\alpha}(\mathbf{k}+\mathbf{K}^\prime)\right].
\end{eqnarray} It is also important to note that for Euclidean action $R$ and $R^\dagger$ (similarly $L$ and $L^\dagger$) are two independent Grassmann spinors. It is also possible to combine the $R$ and $L$ into a four component spinor $\psi^T_\alpha=(R^T_a,L^T_{a})$, and write the effective action as
\begin{eqnarray}
&&S=S_+ + S_- \nonumber \\
&&=\int d^2x d\tau \psi^\dagger_a \left[\partial_\tau + i v\alpha_j\right]\psi_a,
\end{eqnarray}
where $\alpha_j=\tau_j \otimes \eta_3$ for $j=1,2$ are two anticommuting Dirac matrices in the chiral representation. There are three additional anticommuting Dirac matrices $\alpha_3=\tau_3 \otimes \eta_3$, $\beta=\tau_0 \otimes \eta_1$ and $\beta \gamma_5=\tau_0 \eta_2$. The chirality matrix $\gamma_5=\tau_0 \otimes \eta_3$ commutes with $\alpha_j$'s and anticommutes with $\beta$. The $R$ and $L$ sectors are eigenstates of $\gamma_5$ with eigenvalues $\pm 1$. The continuum Hamiltonian anticommutes with $\alpha_3$, $\beta$ and $\beta \gamma_5$ which signifies an emergent $SU(2)$ chiral symmetry. After defining $\bar{\psi}=\psi^\dagger \beta=\psi^\dagger \Gamma_0$ we obtain the Euclidean action of fermions shown in Eq.~(\ref{s1}) which will be used throughout this paper. 

The free fermion action is invariant under the following discrete symmetry operations:

(i) the time reversal:
\begin{eqnarray}\label{sym1}
&& \psi(t,\mathbf{x}) \to \mathcal{T} \psi(-t, \mathbf{x}), \: \bar{\psi}(t,\mathbf{x}) \to -\bar{\psi}(-t,\mathbf{x}) \mathcal{T}, \nonumber \\
&& \mathcal{T}=i \Gamma_5 \Gamma_1 \otimes \sigma_2\mathcal{K},
\end{eqnarray}
where $\mathcal{K}$ stands for complex conjugation;

(ii) the reflection about the x axis:
\begin{eqnarray}\label{sym2}
&& \psi(t,x,y) \to \mathcal{I}_x \psi(t, x,-y), \: \bar{\psi}(t,x,y) \to \bar{\psi}(t,x,-y) \mathcal{I}_x, \nonumber \\ && \mathcal{I}_x=i \Gamma_2 \Gamma_3 \otimes \sigma_0;
\end{eqnarray}
(iii) the reflection about the y axis:
\begin{eqnarray}\label{sym3}
&&\psi(t,x,y) \to \mathcal{I}_y\psi(t, -x,y), \: \bar{\psi}(t,x,y) \to \bar{\psi}(t,-x,y) \mathcal{I}_y, \nonumber \\
&& \mathcal{I}_y=\Gamma_5 \Gamma_1 \otimes \sigma_0;
\end{eqnarray}
(iv) the inversion through the origin:
\begin{eqnarray}\label{sym4}
&& \psi(t,\mathbf{x}) \to \mathcal{P} \psi(t, -\mathbf{x}), \: \bar{\psi}(t,\mathbf{x}) \to \bar{\psi}(t,-\mathbf{x}) \mathcal{P}, \nonumber \\
&& \mathcal{P}=\mathcal{I}_x\mathcal{I}_y=\Gamma_0 \otimes \sigma_0;
\end{eqnarray}
(v) the lattice translations: $\mathbf{r}_i \to \mathbf{r}_i+ \mathbf{R}$, $\mathbf{R}=n_1 \mathbf{a}_1+ n_2 \mathbf{a}_2$, where $n_1 \in \mathbb{Z} $, $n_2 \in \mathbb{Z}$, and
\begin{eqnarray}\label{translation}
&&\psi(t,\mathbf{x}) \to T \psi(t,\mathbf{x}+\mathbf{R}), \: \bar{\psi}(t,\mathbf{x}) \to \bar{\psi}(t,\mathbf{x}+\mathbf{R}) T, \nonumber \\
&& T= \exp \left (i \frac{2 \pi}{3}(n_1+n_2)\Gamma_5\right)= \exp \left( i (-1)^{n_1+n_2} \frac{2 \pi}{3}\Gamma_5 \right); \nonumber \\
\end{eqnarray}
(vi) the rotation by $\pi/3$ about the origin:
\begin{eqnarray}\label{sym5}
&&\psi(t,\mathbf{x}) \to \mathcal{R} \psi(t, \mathbf{x}^{\prime}), \: \bar{\psi}(t,\mathbf{x}) \to \bar{\psi}(t, \mathbf{x}^{\prime}) \mathcal{R}^\dagger, \nonumber \\
&& \mathcal{R}=\cos \frac{2 \pi}{3} \Gamma_0 - i \sin \frac{2 \pi}{3} \Gamma_5 \Gamma_3.
\end{eqnarray}

In the continuum limit the local moments are described by the following QNL$\sigma$M action
\begin{eqnarray}
\mathcal{S}_n=\frac{1}{2 c g}\int d^2x d\tau \left[c^2 (\partial_x \mathbf{n})^2+(\partial_{\tau}\mathbf{n})^2\right] + iS_B[\mathbf{n_j}].
\end{eqnarray}
The coupling constant $g$ has the dimension of length, and there is an antiferromagnetically ordered phase for $g$ smaller than a critical strength $g_c$. In addition, $S_B[\mathbf{n_j}]$ denotes the underlying Berry phase that vanishes inside the magnetically ordered phase. However, inside the paramagnetic phase the Berry phase does not vanish. It is also important to note the absence of a continuum description of the Berry phase term. 

The coupling between the fermions and the QNL$\sigma$M fields is described by
\begin{equation}
\mathcal{S}_{fn}=g \int d^2x d\tau \left [R^\dagger \tau_3 \boldsymbol \sigma \cdot \mathbf{n} R -L^\dagger \tau_3 \boldsymbol \sigma \cdot \mathbf{n} L\right],
\end{equation}
which in the four component notation becomes
\begin{equation}
\mathcal{S}_{fn}=g \int d^2x d\tau \psi^\dagger_{\alpha} \alpha_3 \mathbf{n} \cdot \boldsymbol \sigma_{\alpha \beta} \psi_{\beta}.
\end{equation} Appearance of $\tau_3$ and equivalently $\alpha_3$ matrices represent the breakdown of the inversion or the sublattice symmetry in the presence of AFM order. The chiral symmetry is now reduced from $SU(2)$ to $U(1)$. The matrix $\gamma_5$ is the generator of this $U(1)$ chiral symmetry. This is a continuum version of the discrete translational symmetry of the honeycomb lattice.
It turns out to be more useful to represent the local moment part in terms of another set of Dirac fermions $\chi$ coupled to the collective mode $\mathbf{n}$. For capturing the effects of antiferromagnetic Kondo coupling we have to choose $m_1=-m_2=m$ as in Eq.~(\ref{s2}).

\section{Topological defects of QNL$\sigma$M}\label{defects}
The sigma model in 2+1-dimensions have two important topological defects. There are static nonsingular topological defects called skyrmions, which cost finite energy~\cite{Rajaraman}. After identifying all the points at spatial infinity, the spatial coordinate space $\mathbf{R}^2$ is compactified on the two sphere $S^2$. The skyrmion textures are classified according to the homotopy group $\Pi_2(S^2)=Z$. The explicit form of the skyrmion configurations with topological index $q$, which are also the solutions of the Euler Lagrange equation $\nabla^2 \mathbf{n}=0$ are described by
\begin{eqnarray}
&&\mathbf{n}=\bigg(\frac{2r^q \lambda^q}{r^{2q}+\lambda^{2q}}\cos q\phi, \; \frac{2r^q \lambda^q}{r^{2q}+\lambda^{2q}}\sin q\phi, \; \frac{r^{2q}-\lambda^{2q}}{r^{2q}+\lambda^{2q}}\bigg), \nonumber \\ \\
&&W[\mathbf{n}]=\frac{1}{8\pi} \int dx d\tau \; \epsilon_{\alpha \beta \lambda} \; \epsilon_{ij} \; n_{\alpha} \partial_i n_{\beta} \partial_j n_{\lambda}= q,
\end{eqnarray}
where $\phi=\arctan (x_2/x_1)$. In Fig.~\ref{fig1} we have illustrated a unit skyrmion configuration for the nonlinear sigma model field. Physically the skyrmion density is tied to the underlying scalar spin chirality. Inside the magnetically ordered phase, the conserved skyrmion current density is defined as
\begin{eqnarray}
j_{\mu,sk}=\frac{\epsilon_{\mu \nu \lambda}} {4\pi} \mathbf{n} \cdot \left( \partial_\nu \mathbf{n} \times \partial_\lambda \mathbf{n}\right).
\end{eqnarray} Therefore, the conserved skyrmion number or charge is well defined and equals to the Pontryagin index
\begin{equation}
Q_{sk}=\int d^2x j_{0,sk}=W_{sk}[\mathbf{n}].
\end{equation} 
In the $CP^{1}$ formulation, one can give skyrmions more physically appealing interpretation. Within this formalism, one introduces a complex, two component bosonic spinor $\mathbf{z}$ satisfying the constraint $\mathbf{z}^\dagger \mathbf{z}=1$, and defines O(3) field as $\mathbf{n}=\mathbf{z}^\dagger \boldsymbol \sigma \mathbf{z}$. A skyrmion in the $CP^1$ formulation describes the presence of gauge flux (``magnetic flux") $2\pi q$ through the $xy$ plane.

The energy cost for a skyrmion is proportional to the stiffness of the sigma model and the topological charge, and it is given by
\begin{equation}
E_{sk}=2\pi \rho_s W_{sk}[\mathbf{n}].
\end{equation} Therefore, topologically distinct skyrmion configurations inside the magnetically ordered state (Higgs phase of the $CP^1$ model) are energetically non-degenerate. When the magnetic phase is destroyed by quantum fluctuations beyond a critical coupling $g_c$, the spin stiffness vanishes. Consequently, the skyrmion excitation gap also vanishes and all the topologically distinct ground states labeled by the skyrmion number become energetically degenerate. This is very similar to what goes on for non-Abelian gauge theory in (3+1)-dimensions. The topologically distinct pure gauge configurations of a non-Abelian gauge field cause enormous ground state (vacuum) degeneracy. For non-Abelian gauge theories, such ground state degeneracy is lifted by the tunneling events between topologically distinct pure gauge configurations or instantons~\cite{Shuryak,Diakonov}. A similar phenomenon can also occur in the paramagnetic phase of (2+1)-dimensional O(3) nonlinear sigma model.

The tunneling between two states with different skyrmion numbers can occur through a singular hedgehog configuration in the Euclidean space-time~\cite{Duncan,Read}. Within the $CP^1$ formulation the hedgehogs are monopoles of the compact U(1) gauge field. Inside the paramagnetic phase, the hedgehogs possess effective finite action $S_h$, and $ e^{-S_h}$ describes the tunneling probability. These tunneling singularities are also classified according to homotopy relation $\Pi_2(S^2)=Z$, but it involves the mapping of a sphere surrounding the singularity onto the order parameter space (another sphere)~\cite{Arafune}. The topological invariant of the hedgehog is given by
\begin{equation}
q_h=\frac{1}{8\pi} \int d^2S_{a} \epsilon_{abc} \; \epsilon_{\alpha \beta \lambda} \; n_{\alpha} \partial_b n_{\beta} \partial_c n_{\lambda},
\end{equation}
where the integral is performed over a sphere surrounding the singularity. The $q_h= \pm 1$ radial (anti)hedgehog corresponds to $\mathbf{n}= \pm x_{\mu}/x$, or any version of them obtained after applying $O(3)$ rotations. By applying Gauss's law we can write 
\begin{eqnarray}
\int d^3r \partial_\mu j_{\mu,sk}= \int d\hat{S}_\mu j_{\mu,sk}=q_h,\label{inst1}
\end{eqnarray} which explicitly demonstrates that a hedgegog violates the skyrmion current conservation law. Furthermore, from the above relation we can also show that
\begin{eqnarray}
q_h=W_{sk}[\tau=\infty]-W_{sk}[\tau=-\infty].
\end{eqnarray}

These dynamic singular configurations are responsible for giving rise to the Berry's phase~\cite{Duncan,Read}. In the ordered phase, the hedgehog and anti-hedgehog are linearly confined, and the Berry's phase vanishes. Only in the paramagnetic phase, the hedgehogs can give rise to a nontrivial Berry's phase. In contrast to the (1+1) dimensions, we do not have a continuum description for $S_B[\mathbf{n}]$ in (2+1) dimensions. Only for a simultaneously time reversal and parity (spatial reflection symmetry) breaking theory, we can have a continuum description of the Berry's phase as a topological theta or a Hopf term (which in the $CP^1$ formalism arises as a Chern Simons term for the U(1) gauge fields). In the absence of $P$ breaking, the Berry's phase depends on the size of the spin $S$ and the lattice coordination number $Z$ according to the formula
\begin{equation}
S_B[\mathbf{n}]=\int d\tau \sum_j \frac{4S \pi}{Z} \xi_j q_{h,j},
\end{equation}
where $j$ specifies the dual lattice sites, and $q_{h,j}$ is the topological charge of the hedgehogs located at $j$~\cite{Read}. The dual lattice is partitioned into $Z$ sublattices and the integer valued weight factors $\xi_j=0,1,...,Z-1$ on the different sublattices. Consequently, there is a periodicity $2S (modulo Z)$. On a honeycomb lattice $Z=3$, and the Berry's phase determines the pattern of the $C_{3v}$ symmetry breaking due to the spin Peierls order for different quantized value of the spin. For $2S=0 (modulo 3)$, Berry's phase is absent and there is no spin Peierls order, and the disordered ground state is nondegenerate. When $2S=1 (modulo 3)$, the disordered ground state has threefold degeneracy, and corresponds to the Peierls order.

\begin{table}[htbp]
\caption{The transformation properties of the competing singlet orders under the discrete symmetry operations, for single species of fermions, in the absence of the Kondo coupling. Under the translation $T$, the bilinears $\mathcal{O}_{M,\psi}=\bar{\psi} \hat{M} \otimes \sigma_0 \; e^{i \phi \Gamma_5} \psi \to \bar{\psi} \hat{M} \otimes \sigma_0 \; e^{i (\phi+\frac{4\pi}{3}) \Gamma_5} \psi$. The even and odd properties under the symmetry operations are respectively denoted by $+$ and $-$ signs.}
\begin{center}
\begin{tabular}{|l|l|l|l|l|l|}
\hline
Bilinear & $\mathcal{T}$ & $\mathcal{I}_x$ & $\mathcal{I}_y$ & $\mathcal{P}$ & \: \: \: \: \: \: \: \: \: \: \: $\mathcal{R}$\\
\hline
$\bar{\psi}\psi$& $+$ & $+$ & $+$ & $+$ & \: \: \: \: \: \: \: \: \: \: \: $+$ \\ \hline
$\bar{\psi}i\Gamma_5\psi$ & $+$ & $+$ & $-$ & $-$ & \: \: \: \: \: \: \: \: \: \: \: $-$\\ \hline
$\bar{\psi}\Gamma_{01}\psi$ & $+$ & $+$ & $-$ & $-$ & $-\cos \frac{4\pi}{3}\bar{\psi}\Gamma_{01}\psi+\sin \frac{4\pi}{3}\bar{\psi}\Gamma_{02}\chi$ \\ \hline
$\bar{\psi}i\Gamma_{01}\Gamma_5\psi$ &$+$ & $+$ & $+$ & $+$& $\cos \frac{4\pi}{3}\bar{\psi}i\Gamma_{01}\Gamma_5\psi-\sin \frac{4\pi}{3}\bar{\psi}i\Gamma_{02}\Gamma_5\psi$\\ \hline
$\bar{\psi}\Gamma_{02}\psi$ & $+$ & $-$ & $+$ & $-$ & $-\cos \frac{4\pi}{3}\bar{\psi}\Gamma_{02}\psi-\sin \frac{4\pi}{3}\bar{\psi}\Gamma_{01}\psi$ \\ \hline
$\bar{\psi}i\Gamma_{02}\Gamma_5\psi$ & $+$ & $-$ & $-$ & $+$ &$\cos \frac{4\pi}{3}\bar{\psi}i\Gamma_{02}\Gamma_5\psi+\sin \frac{4\pi}{3}\bar{\psi}i\Gamma_{01}\Gamma_5\psi$ \\ \hline
$\bar{\psi}\Gamma_{03}\psi$ & $-$ & $-$ & $+$ & $-$& \: \: \: \: \: \: \: \: \: \: \: $-$\\ \hline
$\bar{\psi}i\Gamma_{03}\Gamma_5\psi$ & $-$ & $-$ & $-$ & $+$ & \: \: \: \: \: \: \: \: \: \: \: $+$\\ \hline
\end{tabular}
\end{center}
\label{table1}
\end{table}

\section{Induced fermion number of skyrmions} \label{Induced1}
Now we look for the effects of skyrmion excitations on the fermionic sector of the magnetically ordered phase. For a sufficiently large core size, the variation of the sigma model field is weak, and we can perform a gradient expansion calculation~\cite{Wilczek1,Jaroszewicz,Wilczek2,Abanov}. Within the gradient expansion scheme, the skyrmion configurations give rise to the following relations
\begin{eqnarray}
j_{R,\mu}+j_{L,\mu}=\bar{\psi} \Gamma_\mu \psi=0,\\
j_{R,\mu}-j_{L,\mu}=\bar{\psi} \Gamma_\mu \Gamma_5 \psi= 2j_{sk,\mu} \label{induced1}.
\end{eqnarray} The difference between right and left handed fermion currents is also known as the chiral current, and its expectation value is obtained as 
\begin{eqnarray}
&&\langle \bar{\psi} \Gamma_\mu \Gamma_5 \psi \rangle= \mathrm{Tr}\left[\frac{\Gamma_\mu \Gamma_5}{i \Gamma_\rho \partial_\rho+ m \Gamma_3 \mathbf{n} \cdot \boldsymbol \sigma}\right] \nonumber \\
&&=\mathrm{Tr}\left[\frac{\Gamma_\mu \Gamma_5(i \Gamma_\nu \partial_\nu+ m \Gamma_3 \mathbf{n} \cdot \boldsymbol \sigma)}{- \partial^2 + m^2+ i m\Gamma_\rho \Gamma_3 \partial_\rho\mathbf{n} \cdot \boldsymbol \sigma}\right] \nonumber \\
&&=m^3\mathrm{Tr}\bigg[\frac{\Gamma_{\mu}\Gamma_5 \Gamma_{3} \Gamma_{\nu} \Gamma_{3}
\Gamma_{\lambda} \Gamma_3 \mathbf{n}\cdot \boldsymbol \sigma  \partial_{\nu}\mathbf{n}\cdot \boldsymbol
\sigma \partial_{\lambda}\mathbf{n}\cdot \boldsymbol \sigma}{\left(\partial^2+m^2\right)^3}\bigg]
\end{eqnarray}
The trace in the above formula consists of a matrix trace and also integral over the spatial coordinates.
The matrix trace leads to $8 \times \epsilon_{\mu \nu \lambda} \times \epsilon_{abc}$,
and after using the following elementary integral in the energy-momentum space
\begin{equation}
\int \frac{d^3k}{(2\pi)^3} \frac{1}{(k^2+m^2)^3}=\frac{16 \pi}{|m|^3},
\end{equation}
we obtain the result for induced current of Eq.~(\ref{induced1}).
 
Due to the conservation of the skyrmion current inside the magnetically ordered phase, the chiral current is also identically conserved. Consequently, we identify the chiral charge with the skyrmion number
\begin{equation}
q_{ch}=2\int d^2x (j_{R,0}-j_{L,0})=2W_{sk},
\end{equation}
which acts as the generator of the U(1) axial/chiral rotation. This U(1) chiral symmetry is a continuum description of the translational symmetry, and the generator $\Gamma_5$ causes rotation among the chiral symmetry breaking (intervalley) fermion bilinears $\bar{\psi}\psi$ and $\bar{\psi}\Gamma_5 \psi$, and other competing singlets $\bar{\psi}\Gamma_{0j}\psi$ and $\bar{\psi}\Gamma_{0j}\Gamma_5 \psi$ with $j=1,2,3$. These bilinears and their symmetry properties for a honeycomb lattice are shown in Table~\ref{table1}. When the tunneling events destroy the skyrmion number conservation, it also concomitantly destroy the axial current conservation law $\partial_\mu j_{\mu,5}=0$. \emph{This arises even though we do not have a notion of chiral anomaly in (2+1) dimensions}. By using Eq.~\ref{inst1} and Eq.~\ref{induced1}, we arrive at
\begin{eqnarray}
\partial_\mu j_{R,\mu}-\partial_\mu j_{L,\mu}=2 \sum_i q_{h,i}\delta^3(\mathbf{r}-\mathbf{r}_i),\label{conservation}
\end{eqnarray}
demonstrating the violation of chiral symmetry by the hedgehogs.

There is an apparent similarity between Eq.~\ref{conservation} and the Adler-Bell-Jackiw chiral anomaly equation in (3+1)-dimensions. The Adler-Bell-Jackiw formula 
\begin{equation}
\partial_\mu j_{\mu,5}= \partial_\mu[j_{R,\mu}-j_{L,\mu}]=\frac{g^2}{32 \pi^2} Tr[F_{\mu \nu}\tilde{F}_{\mu \nu}],
\end{equation} 
describes the violation of separate number conservation laws for the right and left handed Weyl fermions in (3+1)-dimensions. The trace is taken over the color index of non-abelian gauge fields, and it is absent for abelian gauge fields. The right hand side of this equation describes the instanton density of the non-abelian gauge fields. Despite the technical differences regarding the presence or the absence of axial anomaly in a dimension specific manner, the underlying physical picture of chiral symmetry violation by the tunneling events is quite general. Therefore, for both problems of non-linear sigma model coupled to the (2+1)-dimensional Dirac fermions and (3+1)-dimensional QCD, we anticipate instanton driven breakdown of the U(1) chiral symmetry. Due to this similarity, in the subsequent sections we will closely follow the methodology of QCD$_4$ for addressing the fermion-hedgehog scattering. 

\section{Necessity of fermion zero modes}\label{index}
We will first show why the violation of chiral current conservation as suggested by Eq.~(\ref{conservation}) is tied to the existence of fermion zero modes. In this regard, we consider the Euclidean Dirac operator 
\begin{equation}
\mathcal{D}=\sum_{\mu=0}^{2} \; \partial_\mu \Gamma_\mu\otimes \sigma_0 -i m\Gamma_3 \otimes \boldsymbol \sigma \cdot \mathbf{n}
\end{equation}
in the presence of a single radial hedgehog located at the origin (of space-time) $r=0$. Notice that our Dirac operator is antiHermitian and possesses purely imaginary eigenvalues. If we have defined conjugate Dirac spinor $\bar{\psi}$ by absorbing an additional factor of $i$, the modified dynamic Dirac operator $i \mathcal{D}$ would describe a fictitious Hamiltonian in three dimensions. Since $\mathcal{D}$ does not involve the matrix $\Gamma_5$, it has the following spectral symmetry
\begin{equation}
\{\mathcal{D},\Gamma_5\}=0.
\end{equation}
If $\varphi$ is an eigenstate of $\mathcal{D}$ with eigenvalue $\lambda$, due to the spectral symmetry $\Gamma_5 \varphi$ is also an eigenstate of $\mathcal{D}$ with eigenvalue $-\lambda$. Consequently, the eigenstates with nonzero eigenvalues do not contribute to the expectation value $\langle \bar{\psi} \Gamma_5 \psi \rangle$. Rather $\langle \bar{\psi} \Gamma_5 \psi \rangle$ is entirely determined by the zero modes of $\mathcal{D}$, which are also the eigenstates of $\Gamma_5$. Whether $ \Gamma_5 \varphi  =\pm  \varphi$ determines the chirality or valley index of the zero mode eigenfunction. Thus, 
\begin{equation}
\langle \bar{\psi} \Gamma_5 \psi \rangle=n_+-n_-,\label{index1}
\end{equation} where $n_\pm$ are the number of zero modes with chirality $\pm 1$.

We again consider the expectation value of the chiral current operator $\langle \bar{\psi} \Gamma_\mu \Gamma_5 \psi \rangle $. But, instead of using the plane wave basis as used in the gradient expansion scheme, we will employ the exact eigenstates of the Euclidean Dirac operator for computing this expectation value. Following Jackiw and Rebbi~\cite{JackiwRebbi2} we can write 
\begin{eqnarray}
&&\int \partial_\mu \langle \bar{\psi} \Gamma_\mu \Gamma_5 \psi \rangle = \int \partial_\mu \mathrm{Tr}\left[\sum_n \frac{\varphi_n \varphi^\dagger_n}{\lambda_n}\Gamma_\mu \Gamma_5 \right]\nonumber \\
&&=\int \sum_n \frac{1}{\lambda_n} (\mathrm{Tr}[\partial_\mu \phi_n \phi^\dagger_n \Gamma_\mu \Gamma_5]+\mathrm{Tr}[\phi_n \partial_\mu \phi^\dagger_n \Gamma_\mu \Gamma_5]). \nonumber \\
\end{eqnarray} After using the equation of motion and the invariance of the trace under cyclic permutation we find
\begin{eqnarray}
&&\int \partial_\mu \langle \bar{\psi} \Gamma_\mu \Gamma_5 \psi \rangle =2 \int \sum_n \mathrm{Tr}[\varphi^\dagger_n \Gamma_5 \varphi_n]. 
\end{eqnarray}
Now using Eq.~(\ref{conservation}) and Eq.~(\ref{index1}) we obtain 
\begin{eqnarray}
\int \partial_\mu \langle \bar{\psi} \Gamma_\mu \Gamma_5 \psi \rangle =2(n_+-n_-)=2\int \sum_i q_{h}\delta^3(\mathbf{r})=2 W_h.\label{index2}
\end{eqnarray} Therefore, the mechanism for breaking chiral conservation law is intimately tied to the existence of fermion zero modes for the dynamic Dirac operator $\mathcal{D}$ in the instanton background. This relationship between the number of zero modes and the topological invariant $W_h$ of the background field is known as Callias index theorem\cite{Callias2}. On the mathematical ground, if normalizable zero modes for $\mathcal{D}$ exist, they are protected by the index theorem. But, index theorem by itself does not guarantee the existence of normalizable zero modes. Therefore, in the following section we derive the explicit form of zero modes.

\section{Fermion zero modes}\label{zeromodes}
We need to solve the differential equations
\begin{eqnarray}
&&\Gamma_\mu \partial_\mu \psi -i m (r) \Gamma_3 \otimes \boldsymbol \sigma \cdot \mathbf{n} \psi=0, \\
&&\Gamma_\mu \partial_\mu \bar{\psi} +i m(r) \bar{\psi} \Gamma_3 \otimes \boldsymbol \sigma \cdot \mathbf{n} =0.
\end{eqnarray} For most of our discussion we will consider constant amplitude $m(r)=m$ and a single radial (anti)hedgehog configuration $\mathbf{n}=\pm \hat{r}$. These equations of motion suggest that a hedgehog as seen by the field $\psi$ is perceived as an antihedgehog by the conjugate field $\bar{\psi}$ (recall that $\bar{\psi}$ is an independent Grassmann spinor and not the hermitian conjugate of $\psi$). Since, the antiferromagnetic order parameter does not couple two valleys, we can consider the equations for two valleys separately. For the $\pm$ valleys, the Dirac kernels are respectively given by
\begin{equation}
D_{\pm}=\left[\begin{array}{c c }
\partial_0 \sigma_0 \mp m \mathbf{n} \cdot \boldsymbol \sigma & (\partial_1-i\partial_2)\sigma_0\\
(\partial_1+i\partial_2)\sigma_0 & -\partial_0 \sigma_0 \mp m \mathbf{n} \cdot \boldsymbol \sigma
\end{array} \right].
\end{equation}
For a given valley, we are writing the four component spinor as $\psi=(u_{\uparrow}, u_{\downarrow}, v_{\uparrow}, v_{\downarrow})^T$. For the $+$ valley, $u$ and $v$ respectively denote $A$ and $B$ sublattices, while for $-$ valley they correspond to $B$ and $A$ sublattices. 

For the radial (anti) hedgehog configuration we will employ spherical polar coordinates to write $\mathbf{n}=\pm (\sin \theta \cos \phi, \sin \theta \sin \phi, \cos \theta)$, and look for the zero mode solutions in the s-wave channel. In the s-wave channel, all the derivatives with respect to angular variables $\theta$ and $\phi$ drop out, leading to the following set of equations for the $+$ valley in the presence of a hedgehog,
\begin{eqnarray}
\cos \theta (\partial_r-m)u_{\uparrow}+\sin \theta e^{-i\phi}\left(\partial_r v_{\uparrow}- m u_{\downarrow}\right)&=&0, \nonumber  \\
\cos \theta (\partial_r+m)u_{\downarrow}+\sin \theta e^{-i\phi}\left(\partial_r v_{\downarrow}-m e^{i2\phi}u_{\uparrow}\right)&=&0,\nonumber \\
-\cos \theta (\partial_r+m)v_{\uparrow}+\sin \theta e^{i\phi}\left(\partial_r u_{\uparrow}- m e^{-i2\phi}v_{\downarrow}\right)&=&0, \nonumber \\
-\cos \theta (\partial_r-m)v_{\downarrow}+\sin \theta e^{i\phi}\left(\partial_r u_{\downarrow}-m v_{\uparrow}\right)&=&0. \nonumber \\
\end{eqnarray} 
The normalizable solutions can only appear for $u_{\downarrow}$ and $v_{+,\uparrow}$, with $u_{\downarrow}=-v_{+,\uparrow}$ and $u_{\uparrow}=v_{\downarrow}=0$. Thus we find the following two two-component zero modes for the $+$ valley due to a hedgehog
\begin{equation}
\psi_{R,h,\uparrow}=e^{i\varphi_R} \; f(r)\; \left(\begin{array}{c}
0 \\
1 \end{array} \right ), \psi_{R,h,\downarrow}=- e^{i\varphi_R} \; f(r)\; \left(\begin{array}{c}
1 \\
0 \end{array} \right ),
\end{equation}
where $f(r)=\frac{|m|^{3/2}}{\sqrt{\pi}} \; e^{-m r}$, and $e^{i\varphi_R}$ is an arbitrary global phase factor. We do not find any zero mode for the $-$ valley due to a hedgehog. Therefore, for a unit hedgehog we have $n_+=2$, $n_-=0$, and $W_h=1$, which satisfy the index theorem. 

On the other hand, for an antihedgehog, we do not find any normalizable zero mode for the $+$ valley. But, two normalizable zero modes can be found for the $-$ valley, which are given by
\begin{equation}
\psi_{R,h,\uparrow}=e^{i\varphi_L} \; f(r)\; \left(\begin{array}{c}
0 \\
1 \end{array} \right ), \psi_{R,h,\downarrow}=- e^{i\varphi_L} \; f(r)\; \left(\begin{array}{c}
1 \\
0 \end{array} \right ),
\end{equation}
where $e^{i\varphi_L}$ is another independent global phase factor. For a unit antihedgehog we have $n_+=0$, $n_-=2$, and $W_h=-1$, which again satisfy the index theorem. The situation for the conjugate fields is exactly opposite. For a hedgehog, the conjugate spinor for $-$ valley has two zero modes, while for an antihedgehog the zero modes of conjugate spinor occur for $+$ valley. 

As we have chosen a constant amplitude $m$ for the magnet, the zero  modes are exponentially localized. For a space-time dependent amplitude $m(r)$, we will obtain $f(r) \propto \exp\left[-\int^r_0 dr m(r) \right].$ Thus, a smooth amplitude variation will introduce only quantitative modifications. For the constant $M(r)=m$, we can also calculate rest of the eigenvalues by following Refs.~\onlinecite{JackiwRebbi1,Callias1}. It can be shown that the differential equation for the eigenfunctions of $D^\dagger D$ and $DD^\dagger$ for a given valley are same as the Schr{\"o}dinger equation for a nonrelativistic particle in the Coulomb potential~\cite{Callias1}. Therefore, the nonzero eigenvalues can be obtained from the solution of a well known problem. However, the nonzero eigenvalues only cause a quantitative modification of our main results, and will not be considered in this work.

How about a non-radial hedgehog configuration? A non-radial configuration can be obtained by rotating the radial hedgehog about an arbitrary unit vector $\hat{m}$ by an angle $\phi$ with the help of an SO(3) matrix to obtain
\begin{eqnarray}
&&\hat{n}=\mathcal{R}(\hat{m},\phi) \hat{x} \nonumber \\
&&=(\hat{x} \cdot \hat{m}) \hat{m} + \cos \phi \; [\hat{x}-(\hat{x} \cdot \hat{m})]+\sin \phi \; \hat{m} \times \hat{x},
\end{eqnarray} which describes a general hedgehog configuration with $W_h=1$. This can also be achieved in terms of SU(2) matrices. Since, SU(2) is the universal covering group of SO(3), we can write
$$\mathbf{n} \cdot \boldsymbol \sigma =\mathcal{U}^\dagger \hat{x} \cdot \boldsymbol \sigma \mathcal{U}, $$ where $\mathcal{U}=\pm e^{i \phi/2 \hat{m} \cdot \boldsymbol \sigma}$. The $\pm$ describes the same SO(3) rotation for $\phi$ and $\phi+2\pi$, which is a consequence of 2 to 1 homomorphism. Therefore, zero mode wavefunction for an arbitrary hedgehog can be obtained from the previously found wavefunctions for the radial hedgehogs through the SU(2) rotation $\psi(\hat{m}, \phi)=\mathcal{U}^\dagger \psi(\hat{x})$. Notice, that the zero mode wavefunction changes its sign when $\phi \to \phi + 2\pi$.

In the absence of any source field, the existence of fermion zero mode causes the fermion determinant to vanish. Consequently, the effective action of a hedgehog diverges, leading to a vanishing fugacity or probability of an isolated tunneling event. Therefore, fermion zero modes of the dynamic Dirac operator provide a concrete topological mechanism for suppressing isolated instantons of the O(3) nonlinear sigma model. Within the $CP^{1}$ formulation, this implies the suppression of monopoles, through which the U(1) gauge field can become noncompact or deconfined. 

What can we say about a neutral background with an equal number of hedgehogs and antihedgehogs? This is a very complicated problem, as instantons themselves may be in a liquid, solid or gaseous phase. However, in the close proximity of a magnetic quantum critical point, the diverging correlation length $\xi$ controls the average separation between a pair of hedgehog and antihedgehog (see Fig.~\ref{fig3}), and the instanton denisty is expected to be very small. We will assume that the instanton density in the vicinity of the critical point behaves as
$$n \sim a^{-3} (a/\xi)^x.$$ For determining the exponent $x$ we have to explicitly consider the dynamics of the nonlinear sigma model, and a precise determination of $x$ is a challenging problem. In the absence of fermions, $x$ has been computed by Murthy and Sachdev for a $CP^{N-1}$ model in the large $N$ limit~\cite{MurthySachdev}. Within the large $N$ limit it has been concluded that
$$x_q=2 N \rho_q, \mathrm{with} \; \rho_1 \approx 0.0623, \; \rho_2 \approx 0.1556 .$$ For simplicity, we will leave this exponent unspecified, and work with a dilute instanton gas approximation for obtaining qualitative information in the vicinity of the critical point. \emph{For such a background, the zero modes localized on the two tunneling events will hybridize and split away from the zero eigenvalue. Consequently, we expect a finite fugacity in the paramagnetic phase, which can only vanishes at the critical point as an inverse power of the correlation length}. 

If we introduce Grassmann source fields $\eta$ and $\bar{\eta}$ in the path integral as $\bar{\eta} \psi + \bar{\psi} \eta$, we can compute the effects of fermion zero modes on the correlation functions. \emph{The spinor structure of the zero mode wavefunction will only allow intervalley correlation functions}. Therefore, the hedgehog and the antihedgehog creation operators can be respectively coupled to the operators $\bar{\psi}\dagger_{L}\hat{M} \psi_{R}$ and $\bar{\psi}_{R}\hat{M} \psi_{L}$, where $\hat{M}$ is a hermitian matrix chosen from the following sixteen entries: $\tau_0 \otimes \sigma_0$, $\tau_a \otimes \sigma_0$, $\tau_0 \otimes \sigma_j$, and $\tau_a \otimes \sigma_j$. Among them $\tau_0 \otimes \sigma_0$ and $\tau_a \otimes \sigma_0$ correspond to spin singlet bilinears. We will determine $\hat{M}$ in the following section, where we also provide a heuristic estimation of the chiral symmetry breaking condensate, based on the overlap between two zero modes of opposite chirality. For this, we have to average over random locations and orientations of the hedgehog. Such a procedure will help us to pinpoint the precise nature of the chiral symmetry breaking order parameter or $\hat{M}$.

\section{Nature and size of the chiral condensate}\label{overlap}

For obtaining the overlap between two zero modes localized on widely separated hedgehogs and antihedgehogs, we will need the Fourier transform of the zero mode eigenfunction. The Fourier transform of $f(r)$ is given by 
\begin{eqnarray}
&&f(k)=\frac{|m|^{3/2}}{\sqrt{\pi}} \int^\infty_0 r^2 dr \; \int^{1}_{-1} d(\cos \theta) \; \int^{2\pi}_{0} d\varphi \; e^{-m r} \nonumber \\ && \times e^{i k r \cos \theta}= \frac{4 \sqrt{\pi} |m|^{5/2}}{(k^2+m^2)^2},
\end{eqnarray} where $k=(k_0,\mathbf{k})$ is the three-momentum. For simplicity, we will also set $\varphi_{ch}=\varphi_R-\varphi_L=0$. 

\subsection{Overlap of zero modes}
The overlap between two degenerate zero modes localized around the space-time points $r_1$ and $r_2$ is caused by the kinetic part $\Gamma_\mu \partial_\mu$ of the Dirac operator. The overlap matrix elements are determined as 
\begin{eqnarray}
T_{h-ah}= \int d^3r \; (0, \psi^\dagger_-(r-r_1)\mathcal{U}^\dagger_1) \; \mathcal{D}(r) \; \left(\begin{array}{c}
0 \\
\mathcal{U}_2\psi_-(r-r_2) \end{array} \right), \nonumber \\
T_{ah-h}= \int d^3r \; (\psi^\dagger_+(r-r_1)\mathcal{U}^\dagger_1, 0) \; \mathcal{D}(r) \; \left(\begin{array}{c}
\mathcal{U}_2\psi_+(r-r_2) \\
0 \end{array} \right), \nonumber \\
\end{eqnarray} where $\mathcal{U}_j$s describe the arbitrary orientation of the hedgehog and antihedgehog. We have to average over the orientations by performing an SU(2) group integral, by using the identity 
\begin{equation}
\int d\mathcal{U} \; \mathcal{U}_{ij} \; \mathcal{U}^\dagger_{kl}=\frac{1}{2} \delta_{il} \; \delta_{jk}.
\end{equation}
Since our zero mode wavefunctions display an antisymmetric locking of spin and sublattice indices, the averaging over orientation only allows $\bar{\psi}\psi$ as the emergent chiral symmetry breaking operators. This Dirac mass term describes the emergent spin Peierls order. We are not obtaining the other Dirac mass $\bar{\psi}i\Gamma_5\psi$ due to the specific gauge choice $\varphi_{ch}=0$. The rotationally averaged matrix element can be written as
\begin{eqnarray}
\bar{T}_{h-ah}= \frac{64 \pi}{|m|^3} \; \int \frac{d^3k}{(2\pi)^3} \frac{(i \omega)e^{i k \cdot (r_1-r_2)}}{(1+\frac{k^2}{m^2})^4}.
\end{eqnarray} The spread of the spectrum around the zero eigenvalue is determined by
\begin{eqnarray}
&&\Delta \sim \left[n \; \int d^3 r \bar{T}_{h-ah}\bar{T}_{ah-h}\right]^{1/2} \\
&&=\left[ n \left(\frac{64 \pi}{|m|^3}\right)^2 \int \frac{d^3k}{(2\pi)^3}\frac{k^2}{(1+\frac{k^2}{m^2})^4}\right]^{1/2}
\end{eqnarray} After performing the integral we find
\begin{equation}
\Delta \sim \sqrt{\frac{\pi n}{m}} \sim \frac{1}{\sqrt{m}R^{3/2}},
\end{equation}, where $R$ is the average instanton separation. Therefore the size of the condensate will be
\begin{equation}
\langle \bar{\psi} \psi \rangle \sim \pi \frac{n}{\Delta} \sim \sqrt{n m}
\end{equation} If we associate the amplitude $m$ with the inverse correlation length $\xi$, the condensate vanishes at the critical point according to
\begin{equation}
\langle \bar{\psi} \psi \rangle \sim \xi^{-(x+1)/2} .
\end{equation} Instead if we choose $m$ to describe the bare stiffness, then
\begin{equation}
\langle \bar{\psi} \psi \rangle \sim \xi^{-x/2} .
\end{equation} This last estimate will be in qualitative agreement with the results of Read and Sachdev for SU(N) Heisenberg model. However, we note that there is no \emph{a priori} reason for the agreement between the results of two different models.

We can make another important statement regarding the size of the condensate by following the methodology of QCD. For $\varphi_{ch}=0$, the condensate size is determined by
\begin{equation}
\langle \bar{\psi} \psi \rangle = i \int d^3 x \mathrm{Tr}[G(x,x)].
\end{equation} If we introduce an infinitesimal mass $M \bar{\psi} \psi$ for convenience (as an infra red regulator) of formal manipulations, the fermion propagator in the exact eigenbasis will be given by
\begin{equation}
G(x,y)=-\sum \frac{\phi_n(x) \phi^\dagger_n(y)}{\lambda_n + iM}.
\end{equation} For determining the condensate size in the thermodynamic limit we will take the limit $M \to 0$ at the end of the calculations. After noting that the chiral symmetry $\{ \mathcal{D}, \Gamma_5 \}=0$ implies the existence of a state $\Gamma_5 \phi_n$ with eigenvalue $-\lambda_n$. Therefore, the amplitude of the spin Peierls order will be given by
\begin{eqnarray}
\langle \bar{\psi} \psi \rangle = -i \sum_n \frac{1}{\lambda_n + iM} \nonumber \\
=-\sum_{\lambda_n>0} \frac{2M}{\lambda^2_n +M^2} \nonumber \\
=-\int d\lambda \rho(\lambda) \frac{2M}{\lambda^2+M^2} .
\end{eqnarray} Here $\rho(\lambda)$ is the spectral density for the eigenvalues of the space-time Dirac operator. In the random ensemble, the overlap $\Delta$ between the zero modes localized on the instantons and the antiinstantons causes splitting of the zero modes while giving rise to a continuum of states in the vicinity of the zero eigenvalue. For this reason, the density of states at $\rho(\lambda=0) \neq 0$ and
after taking $M \to 0$ limit, we arrive at
\begin{equation}
\langle \bar{\psi} \psi \rangle=-\pi \rho(\lambda=0).
\end{equation} In the context of QCD this is known as the Banks-Casher relation~\cite{Banks}. The reason behind obtaining the same formula as in QCD for a different model in different dimensionality is the existence of the fermion zero modes in the instanton background. The central message of this relation is the following: \emph{the existence of a finite spectral density of the Dirac operator at zero eigenvalue leads to the chiral symmetry breaking, and the ensuing order parameter density is directly proportional to the spectral density at zero eigenvalue.} Therefore, akin to the QCD problem one can numerically and analytically study the spectrum of a Dirac operator in the presence of hedgehogs to obtain further nonperturbative information. Just based on the dimensional analysis we can infer that $ \rho( \lambda=0) \propto \sqrt{n}, $
where $n$ is the instanton density, which is naturally in agreement with our previous estimation. 

\subsection{Unambiguous choice of spin Peierls order}\label{Peierls}
We can put the appearance of spin Peierls order on a stronger footing through the following calculation. This has been somewhat glossed over during our qualitative discussion of overlap between zero modes. Recall that the hedgehog creation operator can couple to a matrix $\hat{M}$ in the form $\bar{\psi}_L \hat{M} \psi_R$. In the context of QCD, such operators are known as the t'Hooft vertex. For our problem, the t'Hooft vertex for hedgehogs is given by
\begin{eqnarray}
Y_h&=&-\int d^3x_h \int d\mathcal{U} \left[\int d^3x \; \bar{\psi}(\mathbf{x}) \boldsymbol \Gamma \cdot \nabla \phi_h(\mathbf{x}-\mathbf{x}_h)\right] \nonumber \\ &\times& \left[\int d^3y \; \bar{\phi}_h(\mathbf{y}-\mathbf{x}_h) \boldsymbol \Gamma \cdot \nabla \psi(\mathbf{y})\right],
\end{eqnarray}
where we are averaging over the position and orientation of hedgehog respectively denoted by $\mathbf{x}_h$ and $\mathcal{U}$. The zero mode wavefunctions are denoted by $\phi_h$ and $\bar{\phi}_h$, which are left handed and conjugate right handed zero modes respectively. In the frequency-momentum space this can be rewritten as
\begin{eqnarray}
Y_h &=&y_h\int d^3k \int d\mathcal{U} \left[\bar{\psi}(\mathbf{k})\boldsymbol \Gamma \cdot \mathbf{k} \phi_h(\mathbf{k})\right]\left[\bar{\phi}_h(\mathbf{k}) \boldsymbol \Gamma \cdot \mathbf{k} \psi(\mathbf{k})\right] \nonumber \\
&=& y_h \; e^{i\varphi_{ch}}\int d^3k \int d\mathcal{U} \bigg[\bar{R}_{a_1\alpha_1}(\mathbf{k}) \tau^j_{a_1b_1} k_j \mathcal{U}_{\alpha_1 \beta_1} \nonumber \\ &\times& \phi_{h,b_1\beta_1}(\mathbf{k})\bigg]\bigg[\bar{\phi}_{h,a_2 \alpha_2}(\mathbf{k}) \tau^l_{a_2 b_2} k_l \mathcal{U}^{\dagger}_{\alpha_2 \beta_2}L_{b_2\beta_2}(\mathbf{k})\bigg] \nonumber \\
&=&\frac{y_h}{2} e^{i\varphi_{ch}}\int d^3k \; f^2(k) \bigg[\bar{R} _{a_1\alpha_1}L_{b_2\alpha_1}k_j k_l \nonumber \\
&&\times \tau^j_{a_1b_1}\tau^{l}_{a_2b_2}\epsilon_{b_1\beta_1}\epsilon_{a_2\beta_1}\bigg] \nonumber \\
&=&\frac{y_h}{2} e^{i\varphi_{ch}}\int d^3k \; f^2(k) \bar{R}_{a_1b_1}L_{b_2\alpha_1}\bigg[k^2 \tau_0 \nonumber \\ && +i(\mathbf{k} \times \mathbf{k})\cdot \boldsymbol \tau \bigg]_{a_1b_2} \nonumber \\
&=&\frac{y_h}{2} e^{i\varphi_{ch}}\int d^3k \; f^2(k) k^2 \; \bar{R} \tau_0 \otimes \sigma_0 L
\end{eqnarray}
where $y_h$ is the fugacity of hedgehogs, and $\varphi_{ch}=\varphi_R-\varphi_L$ is the global chiral phase of the zero mode wavefunctions. Similarly for the antihedgehogs we obtain the vertex
\begin{eqnarray}
Y_{ah}=\frac{y_{ah}}{2} e^{-i\varphi_{ch}}\int d^3k \; f^2(k) k^2 \; \bar{L} \tau_0 \otimes \sigma_0 R.
\end{eqnarray}
After accounting for $y_h=y_{ah}$ due to the neutrality of the background, the net t'Hooft vertex becomes
\begin{eqnarray}
&&Y=Y_h+Y_{ah}=\frac{y_{h}}{2} \; \int d^3k \; f^2(k) k^2 \nonumber \\ &\times& \bigg[e^{i\varphi_{ch}}\bar{R} \tau_0 \otimes \sigma_0 L +e^{-i\varphi_{ch}}\bar{L} \tau_0 \otimes \sigma_0 R\bigg]\nonumber \\
&=&\frac{y_{h}}{2} \; \int d^3k \; f^2(k) k^2 \; \bar{\psi} \exp \left[i \varphi_{ch} \Gamma_5 \otimes \sigma_0 \right]\psi
\end{eqnarray}
which corresponds to an energy-momentum dependent, complex Dirac mass term. If we set $\varphi_{ch}=0$, we recover the results of previous subsection. \emph{Therefore, by averaging over hedgehog location and orientation we can unambiguously identify the emergent singlet order as the spin Peierls order for a single flavor of eight component Dirac fermion}. This scheme can be enormously helpful for narrowing down our search for competing orders in several other problems. In this regard, we provide a concrete example of competing spin Peierls and Kondo singlets for two flavors of eight component Dirac fermions.   

\begin{table}[htbp]
\caption{The transformation properties of the Kondo singlet bilinears under the discrete symmetry operations. We are denoting the Kondo bilinears as $\mathcal{O}_{\mu,a}=\bar{\Psi}\Gamma_\mu \otimes \mu_a\Psi$, and $\mathcal{O}_{\mu5,a}=\bar{\Psi}\Gamma_\mu\Gamma_5 \otimes \mu_a\Psi$, where $\mu=0,1,2,3$, and $a=1,2$. The even and odd properties under the symmetry operations are respectively denoted by $+$ and $-$ signs.}
\begin{center}
\begin{tabular}{|l|l|l|l|l|l|l|}
\hline
Bilinear & \: \: $\mathcal{T}$ & $\mathcal{I}_x$ & $\mathcal{I}_y$ & $\mathcal{P}$ & $T$ & \: \: \: \: \: \: \: \: \: $\mathcal{R}$\\
\hline
$\mathcal{O}_{0,a}$& $(-1)^{a-1}$ & $+$ & $+$ & $+$ & $+$ & \: \: \: \: \: \: \: \: \: $+$ \\ \hline
$\mathcal{O}_{05,a}$ & $(-1)^{a}$ & $+$ & $-$ & $-$ & $+$ & \: \: \: \: \: \: \: \: \: $-$\\ \hline
$\mathcal{O}_{3,a}$ & $(-1)^{a-1}$ & $-$ & $+$ & $-$ & $+$& \: \: \: \: \: \: \: \: \: $-$ \\ \hline
$\mathcal{O}_{35,a}$ &$(-1)^{a}$ & $-$ & $-$ & $+$ & $+$ & \: \: \: \: \: \: \: \: \: $+$ \\ \hline
$\mathcal{O}_{1,a}$ & $(-1)^{a}$ & $+$ & $-$ & $-$ & $+$& $-\cos \frac{4\pi}{3} \mathcal{O}_{1,a}+\sin \frac{4\pi}{3} \mathcal{O}_{2,a}$ \\ \hline
$\mathcal{O}_{15,a}$ & $(-1)^{a-1}$ & $+$ & $+$ & $+$ & $+$& $\cos \frac{4\pi}{3} \mathcal{O}_{15,a}-\sin \frac{4\pi}{3} \mathcal{O}_{25,a}$\\ \hline
$\mathcal{O}_{2,a}$ & $(-1)^{a}$ & $-$ & $+$ & $-$ & $+$& $-\cos \frac{4\pi}{3} \mathcal{O}_{2,a}-\sin \frac{4\pi}{3} \mathcal{O}_{1,a}$\\ \hline
$\mathcal{O}_{25,a}$ & $(-1)^{a-1}$ & $-$ & $-$ & $+$ & $+$ & $\cos \frac{4\pi}{3} \mathcal{O}_{25,a}+\sin \frac{4\pi}{3} \mathcal{O}_{15,a}$\\ \hline
\end{tabular}
\end{center}
\label{table2}
\end{table} 

\section{Spin Peierls vs. Kondo singlets}\label{Kondo}
We will be considering two species of eight-component Dirac fermions $\psi$ and $\chi$ coupled to the O(3) model with opposite signs 
\begin{align}
S_2&=&\int d^3x\;  \bar{\psi} \left[\sum_{\mu=0}^{2} \; \Gamma_\mu \partial_\mu - i m \; \Gamma_3 \otimes \boldsymbol \sigma \cdot \mathbf{n} \right]\psi  \nonumber \\
&&+\int d^3x\; \bar{\chi} \left[\sum_{\mu=0}^{2} \; \Gamma_\mu \partial_\mu +i m \; \Gamma_3 \otimes \boldsymbol \sigma \cdot \mathbf{n} \right]\chi.\nonumber 
\end{align}
On the magnetically ordered side, the gradient expansion in the presence of skyrmion texture leads to 
\begin{eqnarray}
&&\bar{\psi} \Gamma_\mu \psi=0, \bar{\chi} \Gamma_\mu \chi=0\\
&&\bar{\psi} \Gamma_\mu \Gamma_5 \psi=-\bar{\chi}\Gamma_\mu \Gamma_5 \chi= 2j_{sk,\mu} \label{induced2}.
\end{eqnarray}
Therefore, the sum and the difference between the total number of two species are conserved. If we work with the sixteen component spinor $\Psi=(\psi, \chi)^T$, we can say $\bar{\Psi} \Gamma_0 \otimes \mu_0 \Psi$ and $\bar{\Psi} \Gamma_0 \otimes \mu_3 \Psi$ are conserved quantities. Since the sum of chiral currents for two species vanishes, we can also state that $\bar{\Psi} \Gamma_0 \Gamma_5\otimes \mu_0 \Psi$ or the net chiral density is a conserved quantity. However, the difference between the chiral currents carried by two species is related to the skyrmion current according to 
$$\bar{\Psi} \Gamma_\rho \Gamma_5\otimes \mu_3 \Psi= 4j_{sk,\rho}, $$
and the skyrmion number acts as the generator of relative chiral rotation between two species. Based on this equation, we have earlier constructed   
$$\bar{\Psi} \Gamma_\rho  \otimes \mu_{1}\Psi, \; \bar{\Psi}\Gamma_\rho \Gamma_5 \otimes \mu_{1}\Psi, \; \bar{\Psi} \Gamma_\rho  \otimes \mu_{2}\Psi, \; \bar{\Psi}\Gamma_\rho \Gamma_5\otimes \mu_{2}\Psi,$$ with $\rho=0,1,2,3$ as inter-species, intravalley, Kondo singlets and
$$\bar{\Psi} \mu_{0/3}\Psi, i\bar{\Psi}\Gamma_5 \otimes \mu_{0,3}\Psi, \bar{\Psi}\Gamma_{0j}\otimes \mu_{0/3}\Psi, i\bar{\Psi}\Gamma_5\Gamma_{0j} \otimes \mu_{0/3}\Psi,$$ as intra-species, intervalley singlets 
as the candidates for competing singlet order. The symmetry properties of Kondo singlet operators on a honeycomb lattice are shown in Table~\ref{table2}. How can we select the appropriate singlet orders from this list? 

The relative chiral conservation law $\partial_{\rho}j_{\rho,ch,-}=0$ will be directly broken by the instantons in the paramagnetic phase for nucleating the singlet order, and the factor of four on the right hand side indicates that we have to consider four fermion zero modes. Notice that a hedgehog seen by the $\psi$ fermion, is perceived as an antihedgehog by the $\chi$ fermion. Therefore, the zero modes of $\psi$ and $\chi$ fermions will have opposite chirality. After solving the differential equations in the presence of (anti)hedgehog, we indeed find the zero modes for $\chi$ fermions to have (positive) negative chirality. We also note that the zero modes for $\chi$ fermions can possess new global chiral phases $\varphi_{\chi,R}$ and $\varphi_{\chi,L}$. Since there are four zero modes the nonvanishing correlation functions will involve four fermion operators. To be specific the effective vertex for hedgehog will have a schematic form $\bar{\psi}_R \hat{M} \psi_L \bar{\chi}_l \chi_R$, which indeed breaks the relative chiral rotation symmetry generated by $\bar{\Psi} \Gamma_0 \Gamma_5\otimes \mu_3 \Psi$. The actual form of the effective interaction and its calculation is quite involved and it is presented in the Appendix. For (i) $\varphi_{ch,-}=2n\pi$ and (ii) $\varphi_{ch,-}=(2n+1)\pi$
the effective interaction acquires the form
\begin{widetext}
\begin{eqnarray}
&&Y= \frac{y_h}{16} \; \cos(\varphi_{ch,-}) \; \int d^3k_1 d^3k_2 \; k^2_1k^2_2f^2(k_1)f^2(k_2)\; \bigg [\bar{\Psi}\mu_0\Psi(\mathbf{k}_1)\bar{\Psi}\mu_0\Psi(\mathbf{k}_2)+\bar{\Psi}i\Gamma_5\mu_0\Psi(\mathbf{k}_1)\bar{\Psi}i\Gamma_5\mu_0\Psi(\mathbf{k}_2) \nonumber \\ &&+\bar{\Psi}i\mu_1\Gamma_3\Psi(\mathbf{k}_1)\bar{\Psi}i\mu_1\Gamma_3\Psi(\mathbf{k}_2) +\bar{\Psi}i\mu_2\Gamma_3\Psi(\mathbf{k}_1)\bar{\Psi}i\mu_2\Gamma_3\Psi(\mathbf{k}_2) -\bar{\Psi}\mu_3\Psi(\mathbf{k}_1)\bar{\Psi}\mu_3\Psi(\mathbf{k}_2) -\bar{\Psi}i\mu_3\Gamma_5\Psi(\mathbf{k}_1)\bar{\Psi}i\mu_3\Gamma_5\Psi(\mathbf{k}_2) \nonumber \\ &&-\bar{\Psi}\mu_1\Gamma_{35}\Psi(\mathbf{k}_1)\bar{\Psi}\mu_1\Gamma_{35}\Psi(\mathbf{k}_2)-\bar{\Psi}i\mu_2\Gamma_{35}\Psi(\mathbf{k}_1)\bar{\Psi}i\mu_2\Gamma_{35}\Psi(\mathbf{k}_2)\bigg].
\end{eqnarray}
\end{widetext}
This form of interaction is reminiscent of umklapp interactions for spin 1/2 chain, which breaks the $U(1)$ chiral symmetry of Dirac fermions down to $Z_2$, and the sign of umklapp term determines whether an Ising Neel or a spin Peierls phase is nucleated. For the chiral gauge choice $\varphi_{ch,-}=0$, we can nucleate $\bar{\Psi}\mu_0\Psi$, $\bar{\Psi}i\Gamma_5\mu_0\Psi$ as spin Peierls components, and $\bar{\Psi}i\mu_1\Gamma_3\Psi$, $\bar{\Psi}i\mu_2\Gamma_3\Psi$ as inversion symmetry breaking (sublattice staggered) Kondo singlets. Notice that within the zero mode subspace, all four terms have equal strength of interactions, leading to a very strong competition between two types of singlets. Such a strong competition can even give rise to a liquid phase, where different orders exist only at short distance. We are implying that $\bar{\Psi}\mu_0\Psi$, $\bar{\Psi}i\Gamma_5\mu_0\Psi$, $\bar{\Psi}i\mu_1\Gamma_3\Psi$, $\bar{\Psi}i\mu_2\Gamma_3\Psi$ mutually anticommute and form an SO(4) order parameter, which only has an amplitude but lacks the stiffness. Whether such a situation is indeed realized for a microscopic model, or nonzero mode contributions cause a mild breaking of SO(4) symmetry will be investigated in a separate work. In the following section we outline some additional applications of the instanton calculations for itinerant systems. 


\section{Further applications of our approach}
\label{applications} 
One important aspect of the dynamic fermion zero modes is the topological mechanism for instanton suppression.
For the two species case we have found a strong competition between spin Peierls and Kondo singlet orders. When residual interactions from lattice scale or non-zero modes can lift the degeneracy between these two channels, we can have spin-Peierls to Kondo singlet, antiferromagnet to Kondo singlet and antiferromagnet to spin Peierls transitions. For both antiferromagnet to Kondo singlet and antiferromagnet to spin Peierls transitions, our analysis suggests unconventional criticality with suppressed tunneling. In an earlier work~\cite{GoswamiSi2}, we have shown that such transitions may be accompanied by a level 2 WZW term.

This is in line with the case of a single species of eight component Dirac fermions, for which we have also demonstrated that the paramagnetic phase supports spin Peierls order (out of several possible singlet orders), and the expectation value of this order parameter vanishes with a power law at the magnetic critical point. 
This suggests an unconventional critical point between two different ordered states due to suppression of tunneling and also an unified description where both antiferromagnetic and spin Peierls orders should be treated on an equal footing as
\begin{eqnarray}
S^\prime_1=\int d^3x \; \bigg [\sum_{\mu=0}^{2} \; \bar{\psi}\Gamma_\mu \otimes \sigma_0 \partial_\mu \psi + i m \bar{\psi} \Gamma_3 \otimes \boldsymbol \sigma \cdot \mathbf{n} \psi \nonumber \\ + m^\prime_1 \bar{\psi} e^{i \theta \Gamma_5} \psi \bigg].
\end{eqnarray} 
We can combine two different order parameters in terms of an O(5) nonlinear sigma model with an overall amplitude or coupling constant $\sqrt{m^2+(m^\prime_1)^2}$. Interestingly, after integrating out the fermion fields, one obtains an O(5) nonlinear sigma model augmented by a level one topological  Wess-Zumino-Witten (WZW) term~\cite{Abanov,FisherSenthil,TanakaHu}. Thus our dynamic zero mode calculations are providing some insight into the nature of unconventional critical point arising due to the WZW term.   
Our instanton based calculations also suggest a relation between the level of Wess-Zumino-Witten theory and the number of fermion zero modes (two vs. four). Additional work along this line is required for a better understanding of the critical theories.

The methodology developed in this paper can be applied for computing the fermion determinant when $O(d+1)$ nonlinear sigma model is coupled to massless Dirac fermions in $(d+1)$ dimensional space-time. This provides a direct way to obtain a competing O(2) order parameter, which generally anticommutes with the vector order described by the nonlinear sigma model. Therefore, we suspect this methodology to be relevant for understanding the $O(d+3)$ nonlinear sigma model with a Wess-Zumino-Witten term~\cite{Abanov}. An application along this line will be to replace the antiferromagnetic order by a quantum spin Hall order parameter in (2+1) dimensions~\cite{Grover,ChamonRyu,Herbut1,Moon,Chakravarty1}, which couples to the Dirac fermions as $\bar{\psi} \Gamma_{3}\Gamma_5 \otimes \boldsymbol \sigma  \cdot \mathbf{n} \psi$. For such a model, a hedgehog defect gives rise to zero modes for both valleys. By contrast, the antihedgehog leads to zero modes for conjugate spinor fields. By extending our calculation on the quantum disordered side within a dilute gas approximation, we can immediately show the emergence of a superconducting order parameter. However, we emphasize the superconducting mass term will have a frequency-momentum dependent form factor. In (3+1)-dimensions, antiferromagnetic hedgehogs are static defects and zero modes are actually zero energy eigenstates~\cite{Teo,Hosur}. Similarly spin Peierls order parameter is also a three component vector and can support zero energy states. By performing a calculation along the lines of this paper, we can show the nucleation of competing orders and justify an O(6) theory with WZW term~\cite{Hosur}.

This method will also be useful for providing nonperturbative insight into various quenched disorder problems involving Dirac fermions. Particularly in the context of disordered three dimensional topological insulators, superconductors as well as Dirac semimetals, three dimensional Dirac fermions are coupled to different types of replica or supersymmetric nonlinear sigma models (describing diffusons and cooperons). For such models, the fermion zero modes actually represent physical zero energy states with nontrivial multifractal properties. How they provide a nonperturbative framework for obtaining topological invariant of a disordered insulating state and also their relationship to a direct quantum phase transitions between two topologically distinct insulating states~\cite{RyuFurusaki,FuKane,Goswamidisorder} will be discussed in a separate publication. Another interesting direction will be to extend our work to a finite density of Dirac fermions, which can stabilize a translational symmetry breaking (FFLO) superconducting state in the paramagnetic phase.

\section{Conclusion}\label{conclusions} 
We have developed a theoretical framework for studying competing singlet orders induced by topological defects of antiferromagnetic order parameter in a (2+1)-dimensional itinerant system. For making progress, we have modeled the itinerant system by Dirac fermions which are strongly coupled to the quantum disordered O(3) nonlinear sigma model. Our main goal was to compute the fermion determinant in the presence of a fluctuating or short range antiferromagnetically ordered background, which was built out of an equal number of hedgehogs and antihedgehogs. The salient points of our work are summarized below.

On the magnetically ordered side, the topological tunneling singularities, also known as the hedgehogs and antihedgehogs are linearly confined, and the skyrmion number conservation gives rise to a continuity equation for the skyrmion current. Based on the perturbative gradient expansion method, we have shown
 that an appropriate fermion current (denoted as chiral current) equals the skyrmion current with a multiplicative factor determined by the number of fermion flavors. This has allowed us to identify an induced chiral fermion number for the skyrmion excitations, which acts as the generator of the U(1) chiral symmetry. For the problem at hand, the chiral symmetry is a continuum description of the underlying discrete translational symmetry on the honeycomb lattice. The identification of the skyrmion number as the generator of chiral symmetry helps us to classify all the relevant competing orders. If the nonlinear sigma model is used to describe the collective mode of an underlying single species of eight component Dirac fermion (due to two valleys, two sublattices and two spin indices), the relevant singlet orders mix two valleys and break translational symmetry. By contrast, for the Kondo-Heiseberg model, the coupling between the itinerant fermion and the antiferromagnetically ordered local moments can lead to translational symmetry preserving (intravalley) Kondo singlet states as additional competing orders.

We have demonstrated the topological mechanism through which the hedgehogs pick out a specific competing order. In the hedgehog background, the dynamic Dirac operator possesses zero modes of definite chirality (valley quantum number), leading to the anomalous violation of chiral current conservation law inside the paramagnetic phase. The emergence of fermion zero mode and its stability is shown to be a consequence of the Callias index theorem. The fermion zero modes cause vanishing fugacity for isolated tunneling events and provide a topological mechanism for suppressing the tunneling events, which may lead to a deconfined quantum critical point.  Based on the zero mode structure, we have identified a similarity between our (2+1)-dimensional model and (3+1)-dimensional QCD. \emph
{For a topologically neutral background field with an equal number of hedgehogs and antihedgehogs, the overlap between localized zero modes of opposite chirality determines the precise form of the emergent singlet order}. In the close vicinity of magnetic critical point, when instantons can be treated within a dilute gas approximations, we have explicitly derived the overlap between the zero modes. \emph{For a single species of four component fermion, we show how the overlap between zero modes unambiguously selects the dynamic complex Dirac mass or spin Peierls order over the other charge and current density wave orders}. At a technical level this has been identified with the 't Hooft interaction vertex discussed in the context of QCD. The size of chiral symmetry breaking spin-Peierls condensate is proportional to the density of states of the fermion zero modes and it follows an analog of the Banks-Casher formula for QCD. For the Kondo-Heisenberg model, there are two species of four component fermions, and the 't Hooft vertex describes a quartic fermion interaction, which captures the ensuing competition between the spin Peierls order and Kondo singlet formation. We have also discussed several other applications of our theoretical methods for interacting as well as disordered problems.

\acknowledgements 
This work has been supported in part by JQI-NSF-PFC and LPS-MPOCMTC (P.G.), and
by the NSF grant DMR-1611392 and the Robert A. Welch Foundation Grant No. C-1411 (Q.S.). 
We also acknowledge the hospitality of the Aspen Center for Physics (the NSF Grant No. 1066293) and 
the Kavli Institute of Theoretical Physics, UCSB (the NSF Grant No. PHY11-25915).

\appendix

\onecolumngrid

\section{Derivation of 't Hooft vertex for $N_f=2$}\label{KondoPeierls}
If there are $N_f$ flavors of Dirac fermions, we have seen there are $2N_f$ two component zero modes of positive chirality (+ valley index) localized on the hedgehogs in addition to $2N_f$ negative chirality (- valley) zero modes localized on the antihedgehogs. Therefore, the hedgehog creation operator will couple to a nonlocal t'Hooft interaction vertex with $2N_f$-fermion operators. We have already seen that for $N_f=1$, the hedgehog is coupled to only two fermions operators, which leads to the Dirac mass (Peierls order). For $N_f=2$ we will derive the relevant four fermion vertex by following Refs.~\onlinecite{Shuryak,Diakonov}. 

The partition function for the hedgehog ensemble can be written as
\begin{eqnarray}
Z=\prod_f \int D \psi_f D \bar{\psi}_f \exp[-\int d^3x \bar{\psi}_f \mathcal{D} \psi_f] [\bar{V}_h(\psi_f,\bar{\psi}_f)]^{N_+}  [\bar{V}_{ah}(\psi_f,\psi^\dagger_f)]^{N_-},
\end{eqnarray} where $N_\pm$ are the number of (anti)hedgehogs, and $f$ is the flavor index. For $f=1,2$ we respectively have $\psi$ and $\chi$ fermions introduced in the context of Kondo-Heisenberg model. In the above equation
\begin{equation}
V_h=\int d^3 x \bar{\psi}_f(x) D \phi_{h,0}(x) \int d^3 y \bar{\phi}_{h,0} D \psi_f(y)
\end{equation} is the product of the overlap between zero mode wavefunction $\phi$ and the field operator $\psi$. Following the notations of QCD$_4$, we can further denote this averaged product as
\begin{equation}
Y_{\pm}=\int d^3 r_{h/ah} \; \int d\mathcal{U}_{h/ah}\prod_f V_{h/ah}[\psi_f,\bar{\psi}_f]
\end{equation} where $r_{h}$ and $\mathcal{U}_h$ respectively denote the location and the orientation of the hedgehog. By performing an auxiliary integral the t'Hooft vertices $Y_{\pm}$ can be exponentiated as
\begin{eqnarray}
Z= \int \frac{d \lambda_+}{2\pi} \int \frac{d \lambda_-}{2\pi} \exp[N_+\left(\log \frac{n_+}{i \lambda_+}-1\right) + (+ \to -)] 
\prod_f \int D \psi_f D \psi^\dagger_f \exp[-\int d^3x \psi^\dagger_f \mathcal{D} \psi_f + i \lambda_+ Y_+ + i \lambda_- Y_-] . \nonumber \\
\end{eqnarray} 
It is more convenient to write these vertices in the frequency-momentum space. Recalling the Fourier transform of the zero mode wavefunction, we can infer that the induced interactions are very short-ranged, and the strength of the interaction will be governed by the instanton density $n$.

For the problem with an antiferromagnetic Kondo coupling, the hedgehog contribution to the t'Hooft vertex is obtained as
\begin{eqnarray}
Y_h&=&y_h\prod_{i} \; \int d^3k_i \; \delta^3(\mathbf{k}_1+\mathbf{k}_3-\mathbf{k}_2-\mathbf{k}_4) \int d\mathcal{U} \; [\bar{\psi}(\mathbf{k}_1)\boldsymbol \Gamma \cdot \mathbf{k}_1 \phi_{h,\psi}(\mathbf{k}_1) ][\bar{\phi}_{h,\psi}(\mathbf{k}_2) \boldsymbol \Gamma \cdot \mathbf{k}_2 \psi(\mathbf{k}_2)][\bar{\chi}(\mathbf{k}_3)\boldsymbol \Gamma \cdot \mathbf{k}_3 \phi_{h,\chi}(\mathbf{k}_3) ] \nonumber \\ & & \times [\bar{\phi}_{h,\chi}(\mathbf{k}_4) \boldsymbol \Gamma \cdot \mathbf{k}_4 \chi(\mathbf{k}_4)] \nonumber \\
&=&y_h e^{i\varphi_{ch,-}}\prod_{i} \; \int d^3k_i \; f(\mathbf{k}_i)\; \delta^3(\mathbf{k}_1+\mathbf{k}_3-\mathbf{k}_2-\mathbf{k}_4) \int d\mathcal{U} \;
[\bar{L}_{a_1\alpha_1}(\mathbf{k}_1)R_{a_2\beta_2}\bar{r}_{a_3\alpha_3}(\mathbf{k}_3)l_{b_4\beta_4}][\tau^{j_1}_{a_1b_1}\tau^{j_2}_{a_2b_2}\tau^{j_3}_{a_3b_3}\tau^{j_4}_{a_4b_4} \nonumber \\
&& \times k_{1,j_1}k_{2,j_2}k_{3,j_3}k_{4,j_4}][\mathcal{U}_{\alpha_1\beta_1}\mathcal{U}^{\dagger}_{\alpha_2\beta_2}\mathcal{U}_{\alpha_3\beta_3}\mathcal{U}^{\dagger}_{\alpha_4\beta_4}] [\epsilon_{b_1\beta_1}\epsilon_{a_2\alpha_2}\epsilon_{b_3\beta_3}\epsilon_{a_4\alpha_4}]
\end{eqnarray}
After using the $SU(2)$ group integral identity~\cite{Creutz}
\begin{eqnarray}
&&\int d\mathcal{U} \mathcal{U}_{\alpha_1\beta_1}\mathcal{U}^{\dagger}_{\alpha_2\beta_2}\mathcal{U}_{\alpha_3\beta_3}\mathcal{U}^{\dagger}_{\alpha_4\beta_4}=\frac{1}{3}\delta_{\alpha_1\beta_2}\delta_{\beta_1\alpha_2}\delta_{\alpha_3\beta_4}\delta_{\beta_3\alpha_4}+\frac{1}{3}\delta_{\alpha_1\beta_4}\delta_{\beta_1\alpha_4}\delta_{\alpha_2\beta_4}\delta_{\beta_2\alpha_3}-\frac{1}{6}\delta_{\alpha_1\beta_2}\delta_{\alpha_3\beta_4}\delta_{\beta_1\alpha_4}\delta_{\alpha_2\beta_3}\nonumber \\
&&-\frac{1}{6}\delta_{\alpha_1\beta_4}\delta_{\beta_2\alpha_3}\delta_{\beta_1\alpha_2}\delta_{\beta_3\alpha_4},
\end{eqnarray}
we find that the t'Hooft vertex has following four parts
\begin{eqnarray}
Y^1_h&=&\frac{y_h}{3}e^{i\varphi_{ch,-}} \; \prod_{i} \; \int d^3k_i \; f(\mathbf{k}_i) \; \delta^3(\mathbf{k}_1+\mathbf{k}_3-\mathbf{k}_2-\mathbf{k}_4) \; \bar{L}_{a_1,\alpha_1}(\mathbf{k}_1)[\mathbf{k}_1\cdot\mathbf{k}_2\tau_0+i(\mathbf{k}_1\times\mathbf{k}_2)\cdot \boldsymbol \tau]_{a_1b_1}R_{b_1\alpha_1}(\mathbf{k}_2) \nonumber \\ &&\times  \bar{r}_{a_2,\alpha_2}(\mathbf{k}_3)[\mathbf{k}_3\cdot\mathbf{k}_4\tau_0+i(\mathbf{k}_3\times\mathbf{k}_4)\cdot \boldsymbol \tau]_{a_2b_2}l_{b_2\alpha_2}(\mathbf{k}_4), \\
Y^2_h&=&\frac{y_h}{3}e^{i\varphi_{ch,-}} \; \prod_{i} \; \int d^3k_i \; f(\mathbf{k}_i) \; \delta^3(\mathbf{k}_1+\mathbf{k}_3-\mathbf{k}_2-\mathbf{k}_4) \; \bar{L}_{a_1,\alpha_1}(\mathbf{k}_1)[\mathbf{k}_1\cdot\mathbf{k}_4\tau_0+i(\mathbf{k}_1\times\mathbf{k}_4)\cdot \boldsymbol \tau]_{a_1b_1}l_{b_1\alpha_1}(\mathbf{k}_4) \nonumber \\ &&\times  \bar{r}_{a_2,\alpha_2}(\mathbf{k}_3)[\mathbf{k}_3\cdot\mathbf{k}_2\tau_0+i(\mathbf{k}_3\times\mathbf{k}_2)\cdot \boldsymbol \tau]_{a_2b_2}R_{b_2\alpha_2}(\mathbf{k}_2), \\
Y^3_h&=&-\frac{y_h}{6}e^{i\varphi_{ch,-}} \; \prod_{i} \; \int d^3k_i \; f(\mathbf{k}_i) \; \delta^3(\mathbf{k}_1+\mathbf{k}_3-\mathbf{k}_2-\mathbf{k}_4) \; \bar{L}_{a_1,\alpha_1}(\mathbf{k}_1)R_{b_2\alpha_1}(\mathbf{k}_2) \bar{r}_{a_3,\alpha_3}(\mathbf{k}_3)l_{b_4\alpha_3}(\mathbf{k}_4)[\mathbf{k}_1\cdot\mathbf{k}_4\tau_0 \nonumber \\ 
&&+i(\mathbf{k}_1\times\mathbf{k}_4)\cdot \boldsymbol \tau]_{a_1b_4}[\mathbf{k}_3\cdot\mathbf{k}_2\tau_0+i(\mathbf{k}_3\times\mathbf{k}_2)\cdot \boldsymbol \tau]_{a_3b_2}, \\
Y^4_h&=&-\frac{y_h}{6}e^{i\varphi_{ch,-}} \; \prod_{i} \; \int d^3k_i \; f(\mathbf{k}_i) \; \delta^3(\mathbf{k}_1+\mathbf{k}_3-\mathbf{k}_2-\mathbf{k}_4) \; \bar{L}_{a_1,\alpha_1}(\mathbf{k}_1)R_{b_2\beta_2}(\mathbf{k}_2) \bar{r}_{a_3,\beta_2}(\mathbf{k}_3)l_{b_4\alpha_1}(\mathbf{k}_4)[\mathbf{k}_1\cdot\mathbf{k}_2\tau_0 \nonumber \\ 
&&+i(\mathbf{k}_1\times\mathbf{k}_2)\cdot \boldsymbol \tau]_{a_1b_2}[\mathbf{k}_3\cdot\mathbf{k}_4\tau_0+i(\mathbf{k}_3\times\mathbf{k}_4)\cdot \boldsymbol \tau]_{a_3b_4}.
\end{eqnarray}
Similarly for antihedhehog we will obtain four quartic interaction terms which are hermitian conjugates of the four terms obtained for hedgehog vertex. The role of these quartic terms can be made more transparent if we contract (i) $\mathbf{k}_1=\mathbf{k}_2$, $\mathbf{k}_3=\mathbf{k}_4$ in $Y^{1}_{h}$ and $Y^{4}_{h}$, and (ii)  $\mathbf{k}_1=\mathbf{k}_4$, $\mathbf{k}_2=\mathbf{k}_3$ in $Y^{2}_{h}$ and $Y^{3}_{h}$. Through this we obtain the form of interactions that can be easily decoupled via Hubbard Stratonovich transformations. The reduced form of the quartic term $Y^{1}_h$ is given by
\begin{eqnarray}
Y^{1}_h&=&\frac{y_h}{3}e^{i\varphi_{ch,-}} \int d^3k_1 d^3k_2 \; [\bar{L}(\mathbf{k}_1)R(\mathbf{k}_1)k^2_1f^2(k_1)][\bar{r}(\mathbf{k}_2)l(\mathbf{k}_2)k^2_2f^2(k_2)].
\end{eqnarray}
The corresponding term for antihedgehog is 
\begin{eqnarray}
Y^{1}_{ah}&=&\frac{y_h}{3}e^{-i\varphi_{ch,-}} \int d^3k_1 d^3k_2 \; [\bar{R}(\mathbf{k}_1)L(\mathbf{k}_1)k^2_1f^2(k_1)][\bar{l}(\mathbf{k}_2)r(\mathbf{k}_2)k^2_2f^2(k_2)]
\end{eqnarray}
For simplicity let us consider the phase choice (i) $\varphi_{ch,-}=2n\pi$ and (ii) $\varphi_{ch,-}=(2n+1)\pi$. After combining $Y^{1}_{h}$ and $Y^{1}_{ah}$ and setting $\varphi_{ch,-}=2n\pi$, we obtain
\begin{eqnarray}
Y^1=\frac{y_h}{24} \int d^3k_1 d^3k_2 \; k^2_1k^2_2f^2(k_1)f^2(k_2)\; \bigg [\bar{\Psi}\Psi(\mathbf{k}_1)\bar{\Psi}\Psi(\mathbf{k}_2)+\bar{\Psi}i\Gamma_5\Psi(\mathbf{k}_1)\bar{\Psi}i\Gamma_5\Psi(\mathbf{k}_2)-\bar{\Psi}\mu_3\Psi(\mathbf{k}_1)\bar{\Psi}\mu_3\Psi(\mathbf{k}_2) \nonumber \\-\bar{\Psi}i\mu_3\Gamma_5\Psi(\mathbf{k}_1)\bar{\Psi}i\mu_3\Gamma_5\Psi(\mathbf{k}_2)\bigg ]
\end{eqnarray}
Notice that $Y^1$ breaks the symmetry with respect to $\varphi_{ch,-}$, and the interaction changes sign for $\varphi_{ch,-}=(2n+1)\pi$. This is reminiscent of the umklapp interaction in one dimension. For  $\varphi_{ch,-}=2n\pi$, $Y^1$ will nucleate spin Peierls bilinears $\bar{\Psi}\Psi$ and $\bar{\Psi}i\Gamma_5\Psi$, and the relative phase between them remains free as a Goldstone mode (this phase will be locked into the three-fold pattern only after accounting for lattice effects by going beyond the linearized theory). After some algebra we find $Y^2_h$ and its antihedgehog counterpart leading to
\begin{eqnarray}
&&Y^2=\frac{y_h}{24} \int d^3k_1 d^3k_2 \; k^2_1k^2_2f^2(k_1)f^2(k_2)\; \bigg [\bar{\Psi}i\mu_1\Gamma_3\Psi(\mathbf{k}_1)\bar{\Psi}i\mu_1\Gamma_3\Psi(\mathbf{k}_2) +\bar{\Psi}i\mu_2\Gamma_3\Psi(\mathbf{k}_1)\bar{\Psi}i\mu_2\Gamma_3\Psi(\mathbf{k}_2) \nonumber \\ &&-\bar{\Psi}\mu_1\Gamma_{35}\Psi(\mathbf{k}_1)\bar{\Psi}\mu_1\Gamma_{35}\Psi(\mathbf{k}_2) -\bar{\Psi}\mu_2\Gamma_{35}\Psi(\mathbf{k}_1)\bar{\Psi}\mu_2\Gamma_{35}\Psi(\mathbf{k}_2)\bigg ],
\end{eqnarray}
for $\varphi_{ch,-}=2n\pi$. Clearly, this part of the interaction will cause formation of inversion symmetry breaking Kondo singlets
denoted by $\bar{\Psi}i\mu_1\Gamma_3\Psi$ and $\bar{\Psi}i\mu_2\Gamma_3\Psi$. In the absence of any internal gauge field for $f$ electrons, the phase for Kondo singlets will represent $U(1)$ Goldstone mode. This mode will be massive if an internal gauge field is considered. 
After some algebra involving the Fierz identity of Pauli matrices the other two quartic terms become $Y^3=Y^2/2$ and $Y^4=Y^1/2$. Therefore, the net reduced form of quartic interaction is given by
\begin{eqnarray}
Y&=& \frac{y_h}{16} \int d^3k_1 d^3k_2 \; k^2_1k^2_2f^2(k_1)f^2(k_2)\; \bigg [\bar{\Psi}\Psi(\mathbf{k}_1)\bar{\Psi}\Psi(\mathbf{k}_2)+\bar{\Psi}i\Gamma_5\Psi(\mathbf{k}_1)\bar{\Psi}i\Gamma_5\Psi(\mathbf{k}_2)+\bar{\Psi}i\mu_1\Gamma_3\Psi(\mathbf{k}_1)\bar{\Psi}i\mu_1\Gamma_3\Psi(\mathbf{k}_2) \nonumber \\ && +\bar{\Psi}i\mu_2\Gamma_3\Psi(\mathbf{k}_1)\bar{\Psi}i\mu_2\Gamma_3\Psi(\mathbf{k}_2) -\bar{\Psi}\mu_3\Psi(\mathbf{k}_1)\bar{\Psi}\mu_3\Psi(\mathbf{k}_2) -\bar{\Psi}i\mu_3\Gamma_5\Psi(\mathbf{k}_1)\bar{\Psi}i\mu_3\Gamma_5\Psi(\mathbf{k}_2) \nonumber \\ &&-\bar{\Psi}\mu_1\Gamma_{35}\Psi(\mathbf{k}_1)\bar{\Psi}\mu_1\Gamma_{35}\Psi(\mathbf{k}_2)-\bar{\Psi}i\mu_2\Gamma_{35}\Psi(\mathbf{k}_1)\bar{\Psi}i\mu_2\Gamma_{35}\Psi(\mathbf{k}_2)\bigg].
\end{eqnarray}
We notice that the spin Peierls and Kondo channels have equal strengths of interaction from the hedgehog induced zero mode subspace. This is indicative of a very strong competition among translational symmetry breaking spin Peierls and translational symmetry preserving but inversion symmetry breaking Kondo singlet orders. The form of effective interaction is reminiscent of the Nambu-Jona-Lasinio model of chiral symmetry breaking derived for $N=2$ flavor QCD$_4$.

\twocolumngrid


\begin{thebibliography}{}

\bibitem{Si-Nature} Q.~Si, S.~Rabello, K.~Ingersent, and J.~Smith, \emph{Locally critical quantum phase transitions in strongly correlated metals}, Nature \textbf{413}, 804 (2001).

\bibitem{Coleman-JPCM} P.~Coleman, C.~P{\'e}pin, Q.~Si, and R.~Ramazashvili, \emph{How do Fermi liquids get heavy and die?},
J.\ Phys.:\ Conden.\ Matt. \textbf{13}, R723-R738 (2001).

\bibitem{Senthiletal}
T. Senthil, M. Vojta and S. Sachdev, \emph{Weak magnetism and non-Fermi liquids near heavy-fermion critical points}, Phys.\ Rev.\ B \textbf{69}, 035111 (2004).

\bibitem{Paschen} S. Paschen, T. L\"{u}hmann, S. Wirth, P. Gegenwart, O. Trovarelli, C. Geibel, F. Steglich, P. Coleman, and Q. Si, \emph{Hall-effect evolution across a heavy-fermion quantum critical point}, Nature \textbf{432}, 881 (2004).

\bibitem{Shishido} H.~Shishido, R.~Settai, H.~Harima, and Y.~\={O}nuki, \emph{A drastic change of the Fermi surface at a critical pressure
in {C}e{R}h{I}n{$_5$}: dHvA study under pressure}, J.\ Phys.\ Soc.\ Jpn.\ {\bf 74}, 1103-1106 (2005).


\bibitem{Si_PhysicaB2006}Q. Si, \emph{Global magnetic phase diagram and local quantum criticality in heavy fermion metals}, Physica B \textbf{378}, 23 (2006); Q. Si, \emph{Quantum criticality and global phase diagram of magnetic heavy fermions}, Phys. Status Solidi B{\bf 247}, 476 (2010).

\bibitem{Lohneysen_rmp}
H.~v.~L\"{o}hneysen, A.~Rosch, M.~Vojta, and P.~W\"{o}lfle, \emph{Fermi-liquid instabilities at magnetic quantum phase transitions}, Rev.~Mod.~Phys. \textbf{79}, 1015 (2007).

\bibitem{SiSteglich} Q.~Si and F.~Steglich, \emph{Heavy Fermions and Quantum Phase Transitions}, Science \textbf{329}, 1161 (2010).

\bibitem{Duncan} F. D. M. Haldane, \emph{O(3) Nonlinear σ Model and the Topological Distinction between Integer- and Half-Integer-Spin Antiferromagnets in Two Dimensions}, Phys. Rev. Lett. \textbf{61}, 1029 (1988).

\bibitem{Chakravarty} S. Chakravarty, B. I. Halperin, and D. R. Nelson, \emph{Two-dimensional quantum Heisenberg antiferromagnet at low temperatures}, Phys. Rev. B \textbf{39}, 2344 (1989).

\bibitem{MurthySachdev} G. Murthy and S. Sachdev, \emph{Action of hedgehog instantons in the disordered phase of the (2 + 1)-dimensional $CP^{N-1}$ model}, Nucl. Phys. B \textbf{344}, 557 (1990).

\bibitem{Read} N. Read and S. Sachdev, \emph{Spin-Peierls, valence-bond solid, and N\'eel ground states of low-dimensional quantum antiferromagnets}, Phys. Rev. B \textbf{42}, 4568 (1990).

\bibitem{Senthiletal2} T. Senthil, A. Vishwanath, L. Balents, S. Sachdev and M. P. A. Fisher, \emph{Deconfined Quantum Critical Points}, Science \textbf{303}, 1490 (2004).

\bibitem{Senthiletal3} T. Senthil, L. Balents, S. Sachdev, A. Vishwanath, and M. P. A. Fisher, \emph{Quantum criticality beyond the Landau-Ginzburg-Wilson paradigm}, Phys. Rev. B \textbf{70}, 144407 (2004).

\bibitem{Sandvik} A. W. Sandvik, \emph{Evidence for Deconfined Quantum Criticality in a Two-Dimensional Heisenberg Model with Four-
Spin Interactions}, Phys. Rev. Lett. \textbf{98}, 227202 (2007).

\bibitem{Kaul}R. K. Kaul, and A. W. Sandvik, \emph{A lattice model for the SU(N) Neel-VBS quantum phase transition at large N}, Phys. Rev. Lett. \textbf{108}, 137201 (2012). 

\bibitem{Chalker1} A. Nahum, P. Serna, J. T. Chalker, M. Ortuño, and A. M. Somoza, \emph{Emergent SO(5) Symmetry at the N\'eel to Valence-Bond-Solid Transition}, Phys. Rev. Lett. \textbf{115}, 267203 (2015). 

\bibitem{Chalker2}A. Nahum, J. T. Chalker, P. Serna, M. Ortuño, and A. M. Somoza, \emph{Deconfined Quantum Criticality, Scaling Violations, and Classical Loop Models}, Phys. Rev. X \textbf{5}, 041048 (2015). 

\bibitem{Wilczek1} J. Goldstone, and F. Wilczek, \emph{Fractional Quantum Numbers on Solitons}, Phys. Rev. Lett. \textbf{47}, 986 (1981).

\bibitem{Jaroszewicz} T. Jaroszewicz, \emph{Induced fermion current in the σ model in (2 + 1) dimensions}, Phys. Lett. B \textbf{146}, 337 (1984); T. Jaroszewicz, \emph{Induced topological terms, spin and statistics in (2 + 1) dimensions}, Phys. Lett. B \textbf{159}, 299 (1985).

\bibitem{Wilczek2} Y. H. Chen, and F. Wilczek, \emph{Induced Quantum Numbers In Some 2+1 Dimensional Models}, Int. J. Mod. Phys. B \textbf{3}, 117 (1989).

\bibitem{Abanov} A. G. Abanov and P. B. Wiegmann, \emph{Theta-terms in nonlinear sigma-models}, Nucl. Phys. B \textbf{570}, 685 (2000).

\bibitem{Hermele} M. Hermele, T. Senthil, and M. P. A. Fisher, \emph{Algebraic spin liquid as the mother of many competing orders}, Phys. Rev. B \textbf{72}, 104404 (2005).

\bibitem{FisherSenthil} T. Senthil and M. P. A. Fisher, \emph{Competing orders, nonlinear sigma models, and topological terms in quantum magnets}, Phys. Rev. B \textbf{74}, 064405 (2006).

\bibitem{TanakaHu} A. Tanaka and X. Hu, \emph{Many-Body Spin Berry Phases Emerging from the $\pi$-Flux State: Competition between Antiferromagnetism and the Valence-Bond-Solid State}, Phys. Rev. Lett. \textbf{95}, 036402 (2005).

\bibitem{FuSachdev} L. Fu, S. Sachdev, C. Xu, \emph{Geometric phases and competing orders in two dimensions}, Phys. Rev. B \textbf{83}, 165123 (2011).

\bibitem{RoyHerbut} I. F. Herbut, C. K. Lu, B. Roy, \emph{Conserved charges of order-parameter textures in Dirac systems}, Phys. Rev. B \textbf{86}, 075101 (2012).

\bibitem{GoswamiSi2} P. Goswami and Q. Si, \emph{Topological defects of N\'eel order and Kondo singlet formation for Kondo-Heisenberg model on a honeycomb lattice}, Phys. Rev. B \textbf{89}, 045124 (2014).

\bibitem{JackiwRebbi1} R. Jackiw and C. Rebbi, \emph{Solitons with fermion number 1/2}, Phys. Rev. D \textbf{13}, 3398 (1976).

\bibitem{Callias1} C. J. Callias, \emph{Spectra of fermions in monopole fields—Exactly soluble models}, Phys. Rev. D \textbf{16}, 3068 (1977).

\bibitem{Callias2} C. J. Callias, \emph{Axial anomalies and index theorems on open spaces}, Commun. Math. Phys. \textbf{62}, 213 (1978).

\bibitem{Tsvelik}A. M. Tsvelik, \emph{Semiclassical solution of one dimensional model of Kondo insulator}, Phys. Rev. Lett. \textbf{72}, 1048 (1994).

\bibitem{GoswamiSi1} P. Goswami and Q. Si, \emph{Effects of the Berry Phase and Instantons in One-Dimensional Kondo-Heisenberg Model}, Phys. Rev. Lett. \textbf{107}, 126404 (2011).

\bibitem{'T Hooft} G. 't Hooft, \emph{Computation of the quantum effects due to a four-dimensional pseudoparticle}, Phys. Rev. D \textbf{14}, 3432 (1976).

\bibitem{JackiwRebbi2} R. Jackiw and C. Rebbi, \emph{Spinor analysis of Yang-Mills theory}, Phys. Rev. D \textbf{16}, 1052 (1977).

\bibitem{Shuryak} T. Schaefer and E. Shuryak, \emph{Instantons in QCD}, Rev. Mod. Phys. \textbf{70}, 328 (1998).

\bibitem{Diakonov} D. Diakonov, \emph{Instantons at work}, Prog. Part. Nucl. Phys. \textbf{51}, 173 (2003). 

\bibitem{Witten} I. Affleck, J. Harvey, and E. Witten, \emph{Instantons and (Super-)Symmetry breaking in 2+1 dimensions}, Nucl. Phys. B \textbf{206}, 413 (1982).

\bibitem{Grover} T. Grover, and T. Senthil,  \emph{Topological Spin Hall States, Charged Skyrmions, and Superconductivity in Two Dimensions}, Phys. Rev. Lett. \textbf{100}, 156804 (2008).

\bibitem{ChamonRyu} S. Ryu, C. Mudry, C.-Y. Hou, and C. Chamon, \emph{Masses in graphenelike two-dimensional electronic systems: Topological defects in order parameters and their fractional exchange statistics}, Phys. Rev. B \textbf{80}, 205319 (2009).

\bibitem{Herbut1}C. K. Lu, and I. F. Herbut, \emph{Zero Modes and Charged Skyrmions in Graphene Bilayer}, Phys. Rev. Lett. \textbf{108}, 266402 (2012).

\bibitem{Moon} E. G. Moon, \emph{Skyrmions with quadratic band touching fermions: A way to achieve charge 4e superconductivity}, Phys. Rev. B \textbf{85}, 245123 (2012).


\bibitem{Chakravarty1} C. H. Hsu, and S. Chakravarty, \emph{Charge-2e skyrmion condensate in a hidden-order state}, Phys. Rev. B 87, 085114 (2013).

\bibitem{Arafune} J. Arafune, P. G. O. Freund, and C. J. Goebel, \emph{Topology of Higgs fields}, J. Math. Phys. \textbf{16}, 433 (1975).

\bibitem{Rajaraman} R. Rajaraman, {\it Solitons and Instantons} (North Holland, 1987).

\bibitem{Banks} T. Banks and A. Casher, \emph{Chiral symmetry breaking in confining theories}, Nucl. Phys. B \textbf{169}, 103 (1980).


\bibitem{Teo} J. C. Y. Teo and C. L. Kane, \emph{Majorana Fermions and Non-Abelian Statistics in Three Dimensions}, Phys. Rev. Lett. \textbf{104}, 046401 (2010).


\bibitem{Hosur} P. Hosur, S. Ryu, and A. Vishwanath, \emph{Chiral Topological Insulators, Superconductors and other competing orders in three dimensions}, Phys. Rev. B, \textbf{81}, 045120 (2010). 

\bibitem{RyuFurusaki} S. Ryu, C. Mudry, H. Obuse, and A. Furusaki, \emph{Z2 topological term, the global anomaly, and the two-dimensional symplectic symmetry class of Anderson localization}, Phys. Rev. Lett. \textbf{99}, 116601 (2007). 

\bibitem{FuKane} L. Fu, and C. L. Kane, \emph{Topology, Delocalization via Average Symmetry and the Symplectic Anderson Transition}, Phys. Rev. Lett. \textbf{109}, 246605 (2012). 

\bibitem{Goswamidisorder}P. Goswami, and S. Chakravarty, \emph{Quantum criticality between topological and band insulators in (3+1)-dimensions}, Phys. Rev. Lett. \textbf{107}, 196803 (2011);  \emph{Superuniversality of topological quantum phase transition and global phase diagram of dirty topological systems in three dimensions}, arXiv:1603.03763  

\bibitem{Creutz} M. Creutz, {\em Quarks, Gluons, and  Lattices} (Cambridge Univ. Press, Cambridge, 1983).

\end{thebibliography}
\end{document}